\documentclass[aps,prb,twocolumn,showpacs,superscriptaddress, groupedaddress]{revtex4-2}
\usepackage[english]{babel}
\RequirePackage[latin1,utf8]{inputenc}
\usepackage{graphicx, color}
\usepackage{hyperref}
\definecolor{link}{rgb}{0.1,0.1,0.9}
\hypersetup{colorlinks=true,linkcolor=link,citecolor=link,urlcolor=link,linktocpage}
\usepackage{epstopdf}
\usepackage{color}
\usepackage{amsmath}
\usepackage{amssymb}
\usepackage{longtable}
\usepackage{mathtools}
\usepackage{float}
\usepackage{subfigure}
\usepackage{natbib} 
\usepackage{natbib}
\setcitestyle{numbers,square}
\begin{document}

\title{Magnetism in quasi-two-dimensional tri-layer La$_{2.1}$Sr$_{1.9}$Mn$_3$O$_{10}$ manganite}

\affiliation
{School of Physical Sciences, Jawaharlal Nehru University, New Delhi 110067, India}
\author{Jeetendra Kumar Tiwari}
\affiliation
{School of Physical Sciences, Jawaharlal Nehru University, New Delhi 110067, India}
\author{Birendra Kumar}
\affiliation
{School of Physical Sciences, Jawaharlal Nehru University, New Delhi 110067, India}
\author{Harish Chandr Chauhan}
\affiliation
{School of Physical Sciences, Jawaharlal Nehru University, New Delhi 110067, India}
\author{Subhasis Ghosh}
 \email{subhasis.ghosh.jnu@gmail.com}
\affiliation{School of Physical Sciences, Jawaharlal Nehru University, New Delhi 110067, India%
}

\begin{abstract}
	
The tri-layer La$_{3-3x}$Sr$_{1+3x}$Mn$_{3}$O$_{10}$ manganites of Ruddlesden-Popper (RP) series are naturally arranged layered structure with alternate stacking of m-MnO$_2$ (m = 3) planes and rock-salt type block layers (La, Sr)$_2$O$_2$ along c-axis. The dimensionality of the RP series manganites depends on the number of perovskite layers and significantly affects the magnetic and transport properties of the system. The tri-layer La$_{2.1}$Sr$_{1.9}$Mn$_{3}$O$_{10}$ shows second-order magnetic phase transition. The critical behavior of phase transition has been studied around the transition temperature (T$_C$) to understand the low dimensional magnetism in tri-layer La$_{2.1}$Sr$_{1.9}$Mn$_{3}$O$_{10}$ of the Ruddlesden-Popper series manganites. We have determined the critical exponents for tri-layer La$_{2.1}$Sr$_{1.9}$Mn$_{3}$O$_{10}$, which belong to the short-range two-dimensional (2D)-Ising universality class. The low dimensional magnetism in tri-layer La$_{2.1}$Sr$_{1.9}$Mn$_{3}$O$_{10}$ manganite is also explained with the help of renormalization group theoretical approach for short-range 2D-Ising systems. It has been shown that the layered structure of tri-layer La$_{2.1}$Sr$_{1.9}$Mn$_{3}$O$_{10}$ results in three different type of interactions intra-planer ($ J_{ab} $), intra-tri-layer ($ J_{c} $) and inter-tri-layer ($ J' $) such that $ J_{ab} $ $ > $ $ J_{c} $ $ >>  $$ J' $ and competition among these give rise to the canted antiferromagnetic spin structure above T$ _{C} $. Based on the similar magnetic interaction in bi-layer manganite, we propose that the tri-layer La$_{2.1}$Sr$_{1.9}$Mn$_{3}$O$_{10}$ should be able to host the skyrmion below T$ _{C} $ due to its strong anisotropy and layered structure. 
 \end{abstract}

\maketitle

\section{Introduction}

\subsection{Physical properties of naturally layered manganites}

The tri-layer La$_{3-3x}$Sr$_{1+3x}$Mn$_{3}$O$_{10}$ manganites are a member of the Ruddlesden-Popper (RP) series (La, Sr)$_{m+1}$Mn$_{m}$O$_{3m+1}$ manganite perovskites, where m = 1, 2, 3... $\infty$\cite{fawcett1998structure}. The RP series manganites are naturally arranged layered structure with alternate stacking of m-MnO$_2$ planes and rock-salt type block layers (La, Sr)$_2$O$_2$ along c-axis\cite{fawcett1998structure}. In the RP series manganites, the dimensionality depends on the number of perovskite layers and significantly affects the magnetic and transport properties of the system. In manganites, the introduction of a divalent atom in place of a trivalent atom causes the coexistence of Mn$ ^{3+} $ and Mn$ ^{4+} $ ion, which alters the bond length of Mn$ - $O due to the Jahn-Teller (JT) effect\cite{dagotto2013nanoscale, o1971molecular, van1932theory, jahn1937stability}. The 3d orbital of Mn splits into two energy levels t$ _{2g} $ and e$ _{g} $ in the presence of crystal field and JT effect. The doping of divalent atom results in some empty e$ _{g} $-orbital energy levels which facilitate hopping of electrons responsible for transport properties of tri-layer La$_{3-3x}$Sr$_{1+3x}$Mn$_{3}$O$_{10}$. The ferromagnetism (FM) and metal to insulator transition in manganites are governed by the hoping of e$ _{g} $-orbital electrons between the adjoining Mn$ ^{3+} $ and Mn$ ^{4+} $ ions through O. This mechanism is called double exchange (DE) interaction\cite{zener1951interaction}. The most studied oxides of the RP series manganites are bi-layer (n = 2) La$_{2-2x}$Sr$_{1+2x}$Mn$_{2}$O$_{7}$ and infinite-layer (n = $\infty$) La$_{1-x}$Sr$_{x}$MnO$_{3}$. In particular the three-dimensional (3D) infinite-layer La$_{1-x}$Sr$_{x}$MnO$_{3}$ is most widely studied manganite perovskite due to their extraordinary thermal, electronic and magnetic properties\cite{paraskevopoulos2000magnetic, hemberger2002structural, urushibara1995insulator, rao1998colossal, szewczyk2000magnetocaloric, coey1999mixed, dagotto2013nanoscale, von1993giant, jin1994thousandfold, tokura1994giant, chmaissem2003structural}. The 3D La$_{1-x}$Sr$_{x}$MnO$_{3}$ manganite perovskites have continuous stacking of perovskite structure. The bi-layer La$_{2-2x}$Sr$_{1+2x}$Mn$_{2}$O$_{7}$ manganites consists of quasi-two-dimensional (Q2D) MnO$_2$ bi-layers separated by an insulating (La, Sr)$_2$O$_2$ layer\cite{kimura1996interplane} and received growing interest due to their intriguing physical properties\cite{von1993giant, jonker1950ferromagnetic, chahara1993magnetoresistance, jin1994thousandfold, rao1997charge, kimura1996interplane, wang2004magnetic, moritomo1996giant, asano1997two, seshadri1997study, hirota2002spin}. Apart from the extraordinary magnetic and transport properties, recently observed skyrmionic-bubbles in manganite perovskites\cite{yu2014biskyrmion, nagai2012formation, yu2017variation, morikawa2015lorentz} triggered the renewed attention of researchers. A magnetic skyrmion is a topological particle having a local whirl of the spins\cite{nagaosa2013topological, fert2013skyrmions, chauhan2019multiple, yu2011near}. A topological skyrmion formation occurs due to the competition among different interactions such as Heisenberg (HI) interaction, Dzyaloshinskii-Moriya (DM) interaction, long-range dipole interaction and anisotropy\cite{yu2014biskyrmion, nagai2012formation, yu2017variation, morikawa2015lorentz, nagaosa2013topological, fert2013skyrmions, chauhan2019multiple, yu2011near}. In non-centrosymmetric magnetic materials, DM and HI interaction are responsible for skyrmion formation\cite{nagaosa2013topological}. On the other hand, in centrosymmetric magnetic materials, long-range dipole interaction and anisotropy have been proposed to be responsible for the formation of the skyrmions\cite{yu2014biskyrmion}. Though, it is not yet fully understood how the absence of DM interaction can give rise to skyrmions in manganites. The tri-layer La$_{3-3x}$Sr$_{1+3x}$Mn$_{3}$O$_{10}$ manganites have Q2D MnO$_2$ tri-layers separated by (La, Sr)$_2$O$_2$ layer as shown in Fig. {\ref{XRD}(b)}. However, there are remarkably few studies on the transport and magnetic properties of tri-layer La$_{3-3x}$Sr$_{1+3x}$Mn$_{3}$O$_{10}$ manganites. There are only two reports on tri-layer La$_{3-3x}$Sr$_{1+3x}$Mn$_{3}$O$_{10}$ manganites\cite{mahesh1996effect, jung1999electrical}. Both the reports contain a very limited discussion about the structural, magnetic and transport properties of La$_{2.1}$Sr$_{1.9}$Mn$_{3}$O$_{10}$. The scarcity of the studies in tri-layer La$_{3-3x}$Sr$_{1+3x}$Mn$_{3}$O$_{10}$ manganites is because of the inherent difficulty in the synthesis of high-quality samples of the tri-layer manganites. The preparation of tri-layer manganite samples is challenging in comparison to the bi-layer La$_{2-2x}$Sr$_{1+2x}$Mn$_{2}$O$_{7}$ and infinite-layer La$_{1-x}$Sr$_{x}$MnO$_{3}$ manganites due to difficulty in achieving the stable phase. A little mismatch in the stoichiometric ratio of the precursors and a little deviation from the required temperature cycle may result in the formation of 3D infinite-layer or Q2D bi-layer manganites perovskite as an impurity in the matrix of tri-layer manganite. Hence, careful synthesis of tri-layer La$_{3-3x}$Sr$_{1+3x}$Mn$_{3}$O$_{10}$ manganite is required to get a high-quality sample without impurity. As it stands presently, the study of tri-layer La$_{3-3x}$Sr$_{1+3x}$Mn$_{3}$O$_{10}$ manganite is important due to the following issues: (i) the magnetic and transport properties are not at all explored rigorously, (ii) the exchange mechanism responsible for the spin-spin interaction for FM is not known and (iii) recently observed skyrmionic-bubbles in manganites\cite{yu2014biskyrmion, nagai2012formation, yu2017variation, morikawa2015lorentz} indicates that the tri-layer La$_{3-3x}$Sr$_{1+3x}$Mn$_{3}$O$_{10}$ may also be a potential candidate for the skyrmion host material. These issues emphasize that a thorough magnetic analysis of the tri-layer La$_{3-3x}$Sr$_{1+3x}$Mn$_{3}$O$_{10}$ is required to establish the basic understanding of the magnetism and the exchange interaction involved in tri-layer manganite. The magnetic and electrical properties are explored with the help of high precision magnetic and electrical measurements. In order to investigate the exchange mechanism responsible for the spin-spin interaction, a detailed critical analysis of the second-order phase transition has been carried out.

\subsection{Theoretical background and methodology}

The non-equilibrium dynamics of the magnetic systems near the magnetic phase transition temperature (T$_C$) have received an increasing attention\cite{halperin1967generalization, halperin1969scaling, halperin1972calculation, halperin1974renormalization, halperin1976renormalization, hohenberg1977theory, mazenko2008nonequilibrium}. At time t = 0, the system is quenched in the vicinity of the T$_C$ from an equilibrium state away from the T$_C$. This results in a critical slow down due to the slow relaxation towards the new equilibrium state of the system driven by the sudden quenching in the vicinity of T$_C$. Generally, the Langevin-type equation is used to define the theoretical models for the critical dynamics of a system, which is governed by the Ginzburg-Landau theory for conserved or non-conserved order parameters\cite{halperin1972calculation, halperin1974renormalization, halperin1976renormalization, hohenberg1977theory, mazenko2008nonequilibrium}. Different universality classes are deduced from different theoretical models, which depend on the associated conservation laws and model parameters n and d (n $-$ spin dimensionality and d $-$ space dimensionality). The critical analysis of magnetic systems gives rise the vital information such as the universality class, behavior of phase transition and spin interaction of the system.

According to the Landau (mean-field) theory, the magnetic free energy F$_M$ of a second-order magnetic system can be expressed as a power series in the order parameter M in the vicinity of the T$_C$ as 
\begin{equation}
F(M, T) = F(0)+\frac{a(T)}{2}M^2 +\frac{b(T)}{4}M^4 \\
+...-\mu_0{HM},
\end{equation}
where a(T) and b(T) are Landau coefficients and $\mu{_0}$, H and M are the vacuum permeability, magnetic field and magnetization, respectively. The equilibrium condition is defined by the minimization of the F$_M$ as $\partial$F(M, T)/$\partial$M = 0. The minimization of the F$_M$ gives the following equation of state for a magnetic system near T$_C$
\begin{equation}
\mu_0{H} = {a(T)}M +{b(T)}M^3.
\label{LT2}
\end{equation}
This mean field approach fails in a critical region around T$_C$ characterized by the Ginzburg criterion\cite{ginzburg1961some}. The critical behavior of a magnetic system which undergoes a second-order phase transition can be investigated in detail by a series of correlated critical exponents\cite{stanley1971introduction}. The divergence of correlation length $\xi = \xi_0|(T-T_C)/T_C|^{-\nu}$ ($\nu$ $ - $ critical exponent) results in the universal scaling laws for the spontaneous magnetization (M$_S$) and the inverse magnetic susceptibility ($\chi_0$(T)) in the vicinity of the second-order phase transition T$_C$. The M$_S$(T) is defined for T $<$ T$_C$ and characterized by the exponent $\beta$. The $\chi_0$(T) is defined for T $>$ T$_C$ and characterized by the exponent $\gamma$. The isothermal magnetization (M-H) at T$_C$ is characterized by the exponent $\delta$. The M$_S$(T) and $\chi_0$(T) show a power law dependence on the reduced temperature $\epsilon$ = (T-T$ _{C} $)/T$ _{C} $ and the critical magnetization depends on $\mu_{0}$H\cite{stanley1971introduction, Fishertheory, stanley1999scaling}. The critical exponents before and after T$ _{C} $ can be given by
\begin{equation}
{M_S(T) \propto (-\epsilon)^\beta}; \  {\epsilon<0},\  {T< T_C},
\label{Ms}
\end{equation}
\begin{equation}
{\chi_0^{-1}(T) \propto (\epsilon)^\gamma}; \  {\epsilon>0}, \  {T>T_C}
\label{chi}
\end{equation}
and
\begin{equation}
{M \propto (\mu_0H)^{1/\delta}; \  {\epsilon = 0}}, \  {T = T_C}.
\label{Del}
\end{equation}
Generally, the critical exponents associated with the M$_S$ and $\chi_0$(T) should follow the Arrott-Noakes equation of state\cite{arrott1967approximate} in the asymptotic region $|\epsilon|<0.1$
\begin{equation}
(\mu_0H/M)^{1/\gamma} = a\epsilon + bM^{1/\beta},
\label{Arrott Plot}
\end{equation}
where a and b are the material constant. Using scaling hypothesis, the magnetic
equation of state, i.e., the relationship among the variables M($\mu_0$H, $\epsilon$), $\mu_0$H and T in the asymptotic critical region is expressed as\cite{stanley1999scaling, kaul1985static}
\begin{equation}
M(\mu_0H,\epsilon) = {\epsilon^{\beta}}f_{\pm}(\mu_0H/\epsilon^{(\beta+\gamma)}),
\label{scal}
\end{equation}
where $f_{+}$ is defined for $T > T_C$ and $f_{-}$ is for $T < T_C$. The magnetic equation of the state emphasizes that if the choice of the values of the critical exponents is correct, then all the M-H curve should collapse onto two separate curves below and above T$_C$ independently. The Eq. (\ref{scal}) can  be rewritten in terms of the renormalized magnetization m [m = M($\mu_0$H, $\epsilon$)${\epsilon}^{-\beta}$] and renormalized field h [h = $\mu_0$H$\epsilon^{-(\beta+\gamma)}$] as
\begin{equation}
m = f_\pm(h).
\end{equation}
Furthermore, according to the statistical theory, the relations between the critical exponents that limit the number of independent variables to two are given as\cite{stanley1971introduction, huang1987statistical, kadanoff1966scaling, widom1965surface}:
\begin{equation}
\alpha + 2\beta + \gamma = 2~~~~~~~~~{(Rushbrooke~scaling~realtion)}~
\label{Rushbrooke}
\end{equation}
and
\begin{equation}
\delta = 1+\frac{\gamma}{\beta}.~~~~~~~~~~~~(Widom~scaling~realtion)~
\label{Widom}
\end{equation}
   In the present study, we have chosen a tri-layer compound La$_{2.1}$Sr$_{1.9}$Mn$_{3}$O$_{10}$ for x = 0.3, (hereafter referred to as TL-LSMO-0.3, where TL $ - $ stand for tri-layer). The low dimensional magnetism in tri-layer TL-LSMO-0.3 is explained by the critical analysis using different methods, which includes Kouvel-Fisher (KF) method, modified Arrott plots (MAPs) scaling and critical isotherm analysis. Further confirmation of the low dimensionality of the magnetism in TL-LSMO-0.3 is obtained by renormalization group theory. We have shown that the layered manganite TL-LSMO-0.3 has special characteristics that cannot be explained by the 3D universality classes.  

\section{Experimental Details}

A high-quality tri-layer La$_{2.1}$Sr$_{1.9}$Mn$_{3}$O$_{10}$ manganite sample was synthesized through the standard solid state reaction technique. The stoichiometric amount of high purity precursors of La$_2$O$_3$, SrCO$_3$ and MnO$_2$ were grounded together to achieve the homogeneous mixture of the sample. The final mixture was then calcined at 1050 $^\circ$C for 48 h and sintered at 1400 $^\circ$C for 72 h after making pallets. The sample was regrounded after each calcination and sintering, the final sintering process was repeated to achieve the single phase. The room temperature crystal structure and phase purity were determined by the powder X-ray diffraction (PXRD) (Rigaku miniflex 600-X-ray diffractometer with Cu-K$_{\alpha}$ radiation) followed by the Rietveld refinement. The sample was found to be a tetragonal (\textit{I4/mmm}) structure with no impurity peak. The temperature and field-dependent high precision magnetic data were collected using a physical property measurement system (PPMS). The temperature-dependent zero-field cooled (ZFC) and field cooled (FC) magnetization data were obtained under a constant magnetic field of 10 mT in the temperature range 5 $ - $ 300 K. First quadrant field-dependent M-H curves were obtained under a varying magnetic field of 0 $ - $ 7 T (field step is; 0 to 500 mT $\rightarrow$ $\Delta${$\mu_0$H} = 20 mT and 500 mT to 7 T $\rightarrow$ $\Delta${$\mu_0$H} = 200 mT) in the temperature range of 90 to 120 K with $\Delta$T = 1 K. The resistivity of TL-LSMO-0.3 was collected in the temperature range of 10 to 300 K by using PPMS.
\begin{figure}[htb!]
	\centering
	\includegraphics[trim=0.1mm 0.1mm 0.2mm 0.2mm,clip,width=\linewidth]{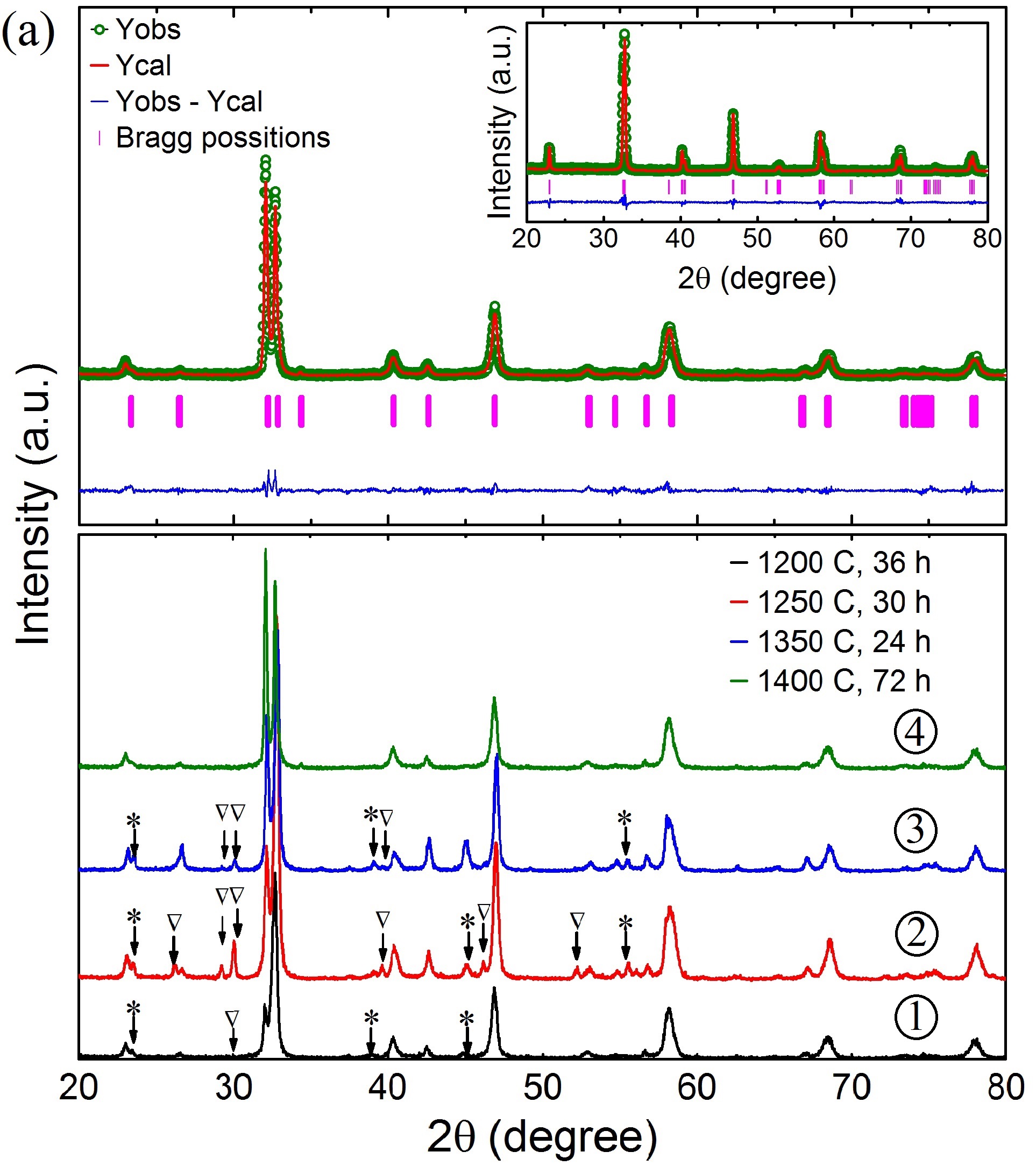}
	\includegraphics[trim=0.1mm 0.1mm 0.2mm 0.2mm,clip,width=\linewidth]{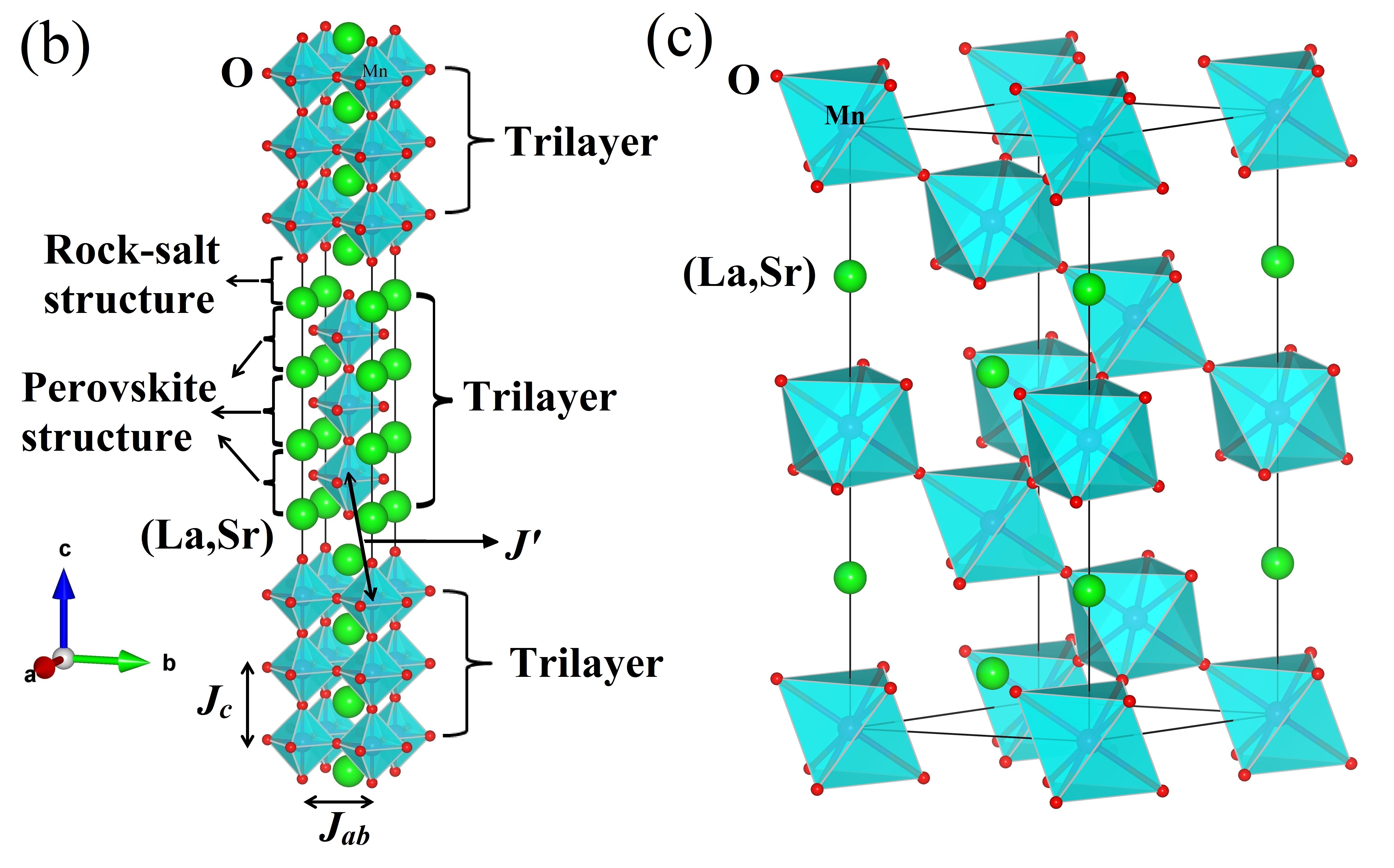}
	\caption{(a) The lower panel shows the PXRD of the tri-layer TL-LSMO-0.3 synthesized under different conditions. The PXRD pattern in green color(pattern 4) is corresponding to the pure phase TL-LSMO-03. Upper panel and inset shows the Rietveld refinement of the single phase TL-LSMO-0.3 and La$ _{0.7} $Sr$ _{0.3} $MnO$_{ 3} $, respectively. (b) and (c) Unit cell crystal structure of the tri-layer La$_{2.1}$Sr$_{1.9}$Mn$_{3}$O$_{10}$ and infinite-layer La$ _{0.7} $Sr$ _{0.3} $MnO$_{ 3} $ manganite. The MnO$_6$ octahedra in the crystalline bulk are denoted in cyan color. Different atoms are shown as the spheres of different  [(La, Sr) - light green, Mn - blue oxygen - red]. The perovskite structure ABO$_3$, [A $\rightarrow$ (La, Sr) and B $\rightarrow$ Mn] consist of a MnO$_6$ octahedra in the center and (La, Sr) atoms lies at the corner of the perovskite structure.}
	\label{XRD}
\end{figure}
The lower panel of the Fig. \ref{XRD}(a) shows the PXRD patterns of samples prepared at different conditions and represented by 1, 2, 3 and 4. The upper panel and inset of Fig. \ref{XRD}(a) shows the Rietveld refinement of the pure phase TL-LSMO-0.3 (pattern 4) and infinite-layer La$ _{0.7} $Sr$ _{0.3} $MnO$_{ 3} $, respectively. Among all the PXRD patterns, the pattern 1, 2 and 3 were not found to be a single phase but showed impurity peaks corresponding to the bi-layer and infinite-layer. The impurity peak formation is either due to the mismatch of stoichiometric ratio and/or the lack of controlled heating rate and heating cycle. The $\star$ and $\triangledown$ symbols correspond to the bi-layer and infinite-layer impurities, respectively. The low dimensionality of the TL-LSMO-0.3 can be understood by comparing the crystal structure of TL-LSMO-0.3 with that of infinite-layer La$ _{0.7} $Sr$ _{0.3} $MnO$_{ 3} $ manganite. Figure \ref{XRD}(b) and (c) show the crystal structure of the TL-LSMO-0.3 and La$ _{0.7} $Sr$ _{0.3} $MnO$_{ 3} $, respectively. The crystal structure of TL-LSMO-0.3 shows layered characteristic in which the rock-salt type structure separates three consecutive perovskite layers and these perovskites layers are made of two-dimensional (2D) network of Mn$ - $O bond and form Q2D MnO$ _{2} $ planes\cite{asano1997two}. In contrast, the crystal structure of infinite-layer La$ _{0.7} $Sr$ _{0.3} $MnO$_{ 3} $ has continuous stacking of the 3D perovskite layers, which is made of a 3D network of Mn$ - $O bond and noticeably different than that of TL-LSMO-0.3\cite{asano1997two}. Hence, the dimensionality of the RP series manganites can be modified by altering the number of perovskite layers\cite{mahesh1996effect}.

\section{Results and analysis}

\begin{figure}[htb!]
	\centering
	\includegraphics[trim=0.1mm 0.1mm 0.2mm 0.2mm,clip,width=\linewidth]{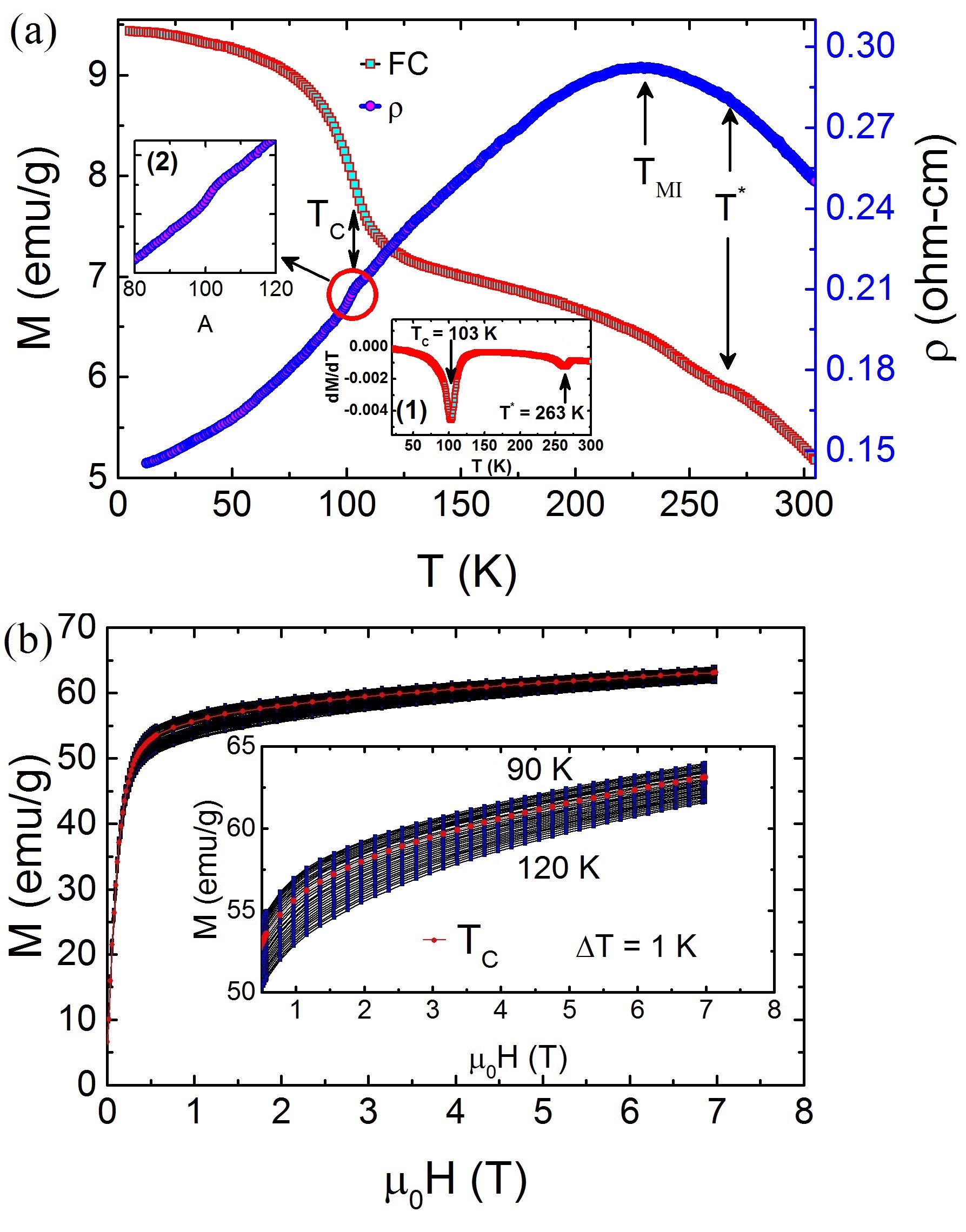}
	\caption{(a) FC of TL-LSMO-0.3 in constant magnetic field of 10 mT. The T$_C$ $\approx103$ K is determined by the minimum of the derivative of magnetization (dM/dT). Resistivity of TL-LSMO-0.3 shows a metal-insulator transition (T$_{MI}$) at $\approx{230}$ K and a step at $\approx{103}$ K corresponding to the magnetic phase transition T$_{C}$ and a anomaly at $\approx{263}$ K represented by T$ ^{*} $. (b) M-H curve for TL-LSMO-0.3 in the temperature range 90 to 120 K with $\Delta$T = 1 K and magnetic field of 7 T.}
	\label{FC_ZFC}
\end{figure}

The FC and resistivity curves for TL-LSMO-0.3 are shown in Fig. \ref{FC_ZFC}(a). The T$_C$ $\approx$ 103 K of the sample is determined by the minimum of the derivative of FC curve, shown in the inset (1) of Fig. \ref{FC_ZFC}(a)\cite{mahesh1996effect, jung1999electrical}. There is another transition T$ ^{*} $ that appears at $\approx$ 263 K in addition to the first transition 103 K. Generally, in the case of infinite-layer manganites, the magnetization above T$ _{C} $ is zero, but in the present sample, the magnetization is non-zero above T$_C$. Similar results have been observed in bi-layer La$_{2-2x}$Sr$_{1+2x}$Mn$_{2}$O$_{7}$ manganites\cite{kimura1996interplane, wang2005magnetic, argyriou1997unconventional}. 
The non-zero magnetization above T$_C$ in bi-layer La$_{2-2x}$Sr$_{1+2x}$Mn$_{2}$O$_{7}$ manganites has been explained by the 2D short-range FM ordering in their PM state\cite{kimura1996interplane, wang2005magnetic, argyriou1997unconventional}. The temperature-dependent resistivity curve for TL-LSMO-0.3 shows a broad peak at $\approx{230}$ K corresponding to the metal-insulator transition (T$_{MI}$) along with a step-like behavior at T$ _{C} $ $\approx{103}$ K [inset (2)] and a small anomaly at $\approx{263}$ K corresponding to the second transition T$ ^{*} $. Generally, in manganites, the T$_{MI}$ of MIT coincides with the T$ _{C} $ of the system. In contrast, the TL-LSMO-0.3 shows a significant difference between the MIT transition T$_{MI}$ and the magnetic transition T$ _{C} $. The metallic behavior of TL-LSMO-0.3 below T$ _{C} $ can be explained by the DE mechanism, where a large number of carriers are available\cite{zener1951interaction, jonker1950ferromagnetic}. On the other hand, the metallic behavior above T$ _{C} $ is due to the formation of FM cluster and can be explained with percolation mechanism, which describes the metallic behavior even in the absence of long-range magnetic ordering. The transition from metallic to insulating phase occurs due to the formation of polarons because of the distortion in MnO$ _{6} $ octahedra\cite{millis1995double}. Figure \ref{FC_ZFC}(b) represents the M-H data of TL-LSMO-0.3 in critical region 90 K $\leq$ T $\leq$ 120 K, where $\Delta$T = 1 K. 

\section{Entropy analysis}

\subsection{Universal curve for second-order phase transition}

\begin{figure}[htb!]
	\centering
	\includegraphics[trim=0.3mm 0.3mm 0.3mm 0.3mm,clip,width=\linewidth]{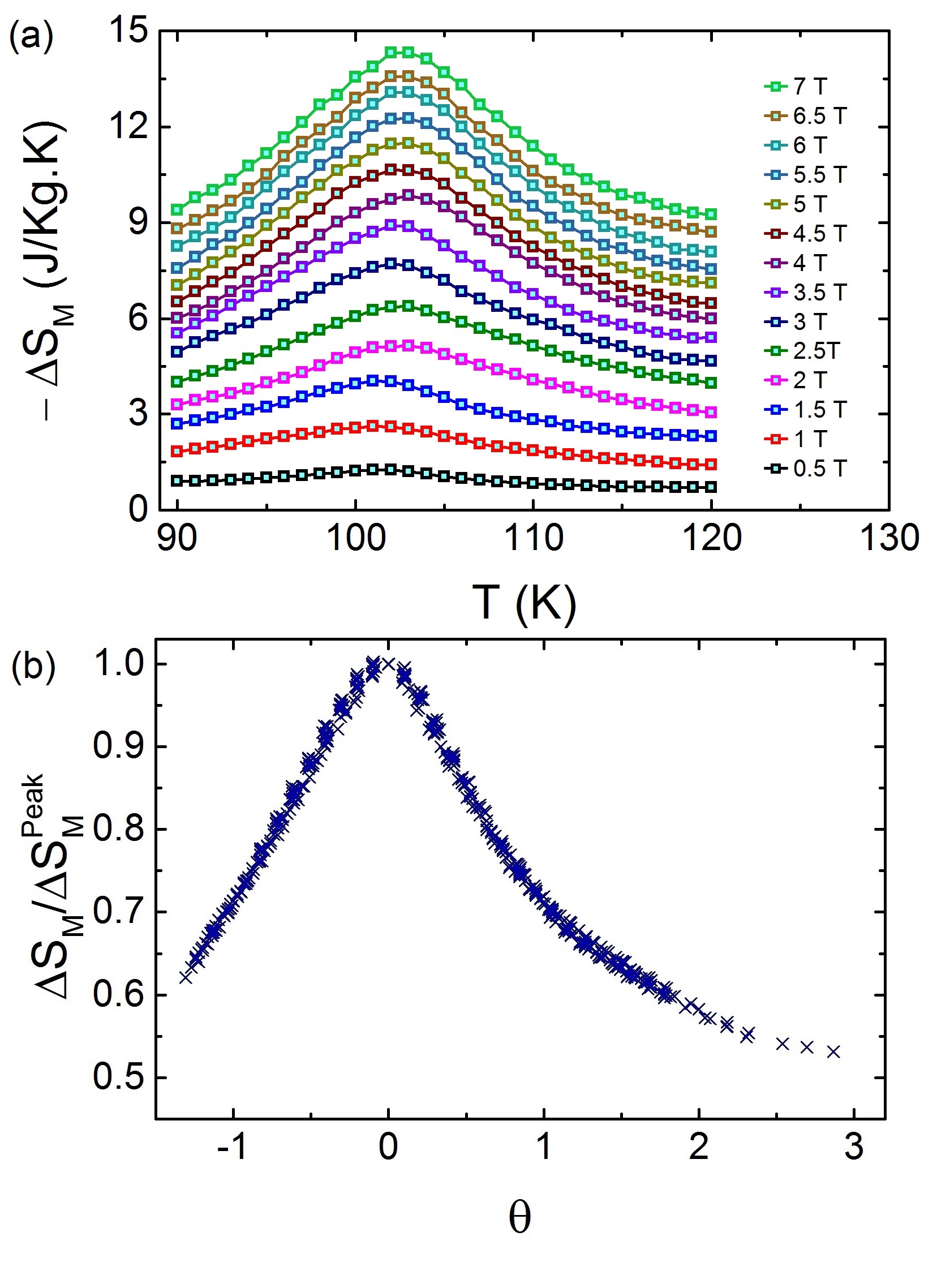}
	\caption{(a) The evolution of $\Delta{S_M}$ vs. T at different fields (0.5 $\rightarrow$ 7 T) determined by M-H curve [Fig. \ref{FC_ZFC}(b)], which shows a continuous nonmonotonic change of $\Delta{S_M}$ around T$_C$. (b) Universal curve for TL-LSMO-0.3 shows the collapse of all $\Delta{S_M}$ on single curve, which is the characteristic of second-order phase transition.}
	\label{Entropy1}
\end{figure}

This section presents systematic study of the behavior of universal curve for magnetic entropy change ({$\Delta{S_M}$}) to confirm the order of the magnetic phase transition in TL-LSMO-0.3. A universal curve should be constructed for field-dependent $\Delta{S_M}$ only in case of second-order phase transition\cite{VFranco, Romero-Muniz, franco2007field, Bonilla, guetari2016structure}. The existence of a universal curve is based on the formalism that the equivalent points of the different {$\Delta{S_M}$} curves calculated for different magnetic fields should collapse on a single curve\cite{VFranco, Romero-Muniz, franco2007field, Bonilla, guetari2016structure}. If TL-LSMO-0.3 shows a second-order magnetic phase transition, all the $\Delta{S_M}$ curves will collapse on a single universal curve. Before starting the analysis of the scaling behavior of {$\Delta{S_M}$}, we have to calculate the temperature variation of {$\Delta{S_M}$}. The {$\Delta{S_M}$} can be calculated by using M-H data and Maxwell's thermodynamic relation given below\cite{Romero-Muniz, bingham2009magnetocaloric}
\begin{eqnarray}
\Delta{S_M}(\mu{_0}H, T) = \Delta{S_M}(\mu{_0}H, T) - \Delta{S_M}(0, T),\nonumber\\
\Delta{S_M}(\mu{_0}H, T) = \int_{0}^{\mu{_0}H_{max}}\Bigg(\frac{\partial{S(\mu{_0}H, T)}}{\partial{(\mu{_0}H)}}\Bigg)_{T}d({\mu{_0}H}).~~~~~
\end{eqnarray}
Using Maxwell's thermodynamic relation
\begin{equation}
\Bigg(\frac{\partial{S(\mu{_0}H, T)}}{\partial{(\mu{_0}H)}}\Bigg)_{T} = \Bigg(\frac{\partial{M}(\mu{_0}H,T)}{\partial{T}}\Bigg)_{\mu_{0}H}.
\end{equation}
Now, above relation {$\Delta{S_M}$} can be expressed as follows
\begin{eqnarray}
\Delta{S_M}(\mu{_0}H, T) = \int_{0}^{\mu{_0}H_{max}}\Bigg(\frac{\partial{M}(\mu{_0}H,T)}{\partial{T}}\Bigg)_{\mu_{0}H}d({\mu{_0}H}).\nonumber\\
\label{Entropy}
\end{eqnarray}
Using isothermal M-H curves, the $\Delta{S_M}$ in the presence of the magnetic field can be calculated numerically from the following equation
\begin{eqnarray}
\Delta{S_M}(\frac{T_{1} + T_{2}}{2}) = \frac{1}{(T_{1} - T_{2})} \Bigg[ \int_{0}^{\mu{_0}H_{max}}\Big({\partial{M}(\mu{_0}H,T_{2}})\nonumber\\ - {\partial{M}(\mu{_0}H,T_{1}}\Big)_{\mu_{0}H}d({\mu{_0}H})\Bigg].~~~~~
\label{Entr}
\end{eqnarray}
Figure \ref{Entropy1}(a) shows the variation of {$\Delta{S_M}$} with temperature and all the curves show a maximum at T$_C$. The value of the {$\Delta{S_M}$} peak increases with magnetic field. In order to construct the universal curve, all the {$\Delta{S_M}$} curves were normalized with their respective maximum entropy change $\Delta{S_M}(T, \mu_{0}H)$/$\Delta{S_M^{peak}}$(T, $\mu_{0}$H). Next, the temperature axis is rescaled by considering the reference temperature such that {$\Delta{S_M}(T_{r})$/$\Delta{S_M^{peak}}$} $ \geq {l}$, where T$_r$ is the reference temperature and $ l $ (0 $ < $ $ l $ $\leq$ 1) is the arbitrary constant. Although, $ l $ can take any value between 0 to 1 but large value of $ l $, i.e., the reference temperature chosen very close to $\Delta{S_M^{peak}}$ may result a large numerical errors due to limited number of points. We define the new rescaled temperature axis ($\theta$) as 
\begin{equation}
	\theta =  \begin{cases}
		-(T-T{_C})/(T_{r_1}-T_{C}), \ T \leq T_{C} \\
		(T-T{_C})/(T_{r_2}-T_{C}), \ T > T_{C},
	\end{cases}
	\label{Entropyscaling}
\end{equation}
where T$_{r_1}$ and T$_{r_2}$ are the two reference temperatures for T $\leq$ T$_{C}$ and T $ > $ T$_{C}$ respectively. The reference temperatures T $\leq$ T$_{C}$ and T $ > $ T$_{C}$ are selected such that $\Delta{S_M(T_{r_1})}$/$\Delta{S_M^{peak}}$ = $\Delta{S_M}(T_{r_2})$/$\Delta{S_M^{peak}}$ = 0.7. The universal curve for TL-LSMO-0.3 is plotted in Fig. \ref{Entropy1}(b), which shows the collapse of all the $\Delta{S_M}$ curves calculated at different field on a single curve. The formation of single curve for TL-LSMO-0.3 confirms the second-order phase transition around T$ _{C} $.

\begin{figure*}[htb!]
	\centering
	\begin{tabular}{cccc}
		\includegraphics[trim=0.1mm 0.1mm 0.1mm 0.1mm,clip,width=85mm]{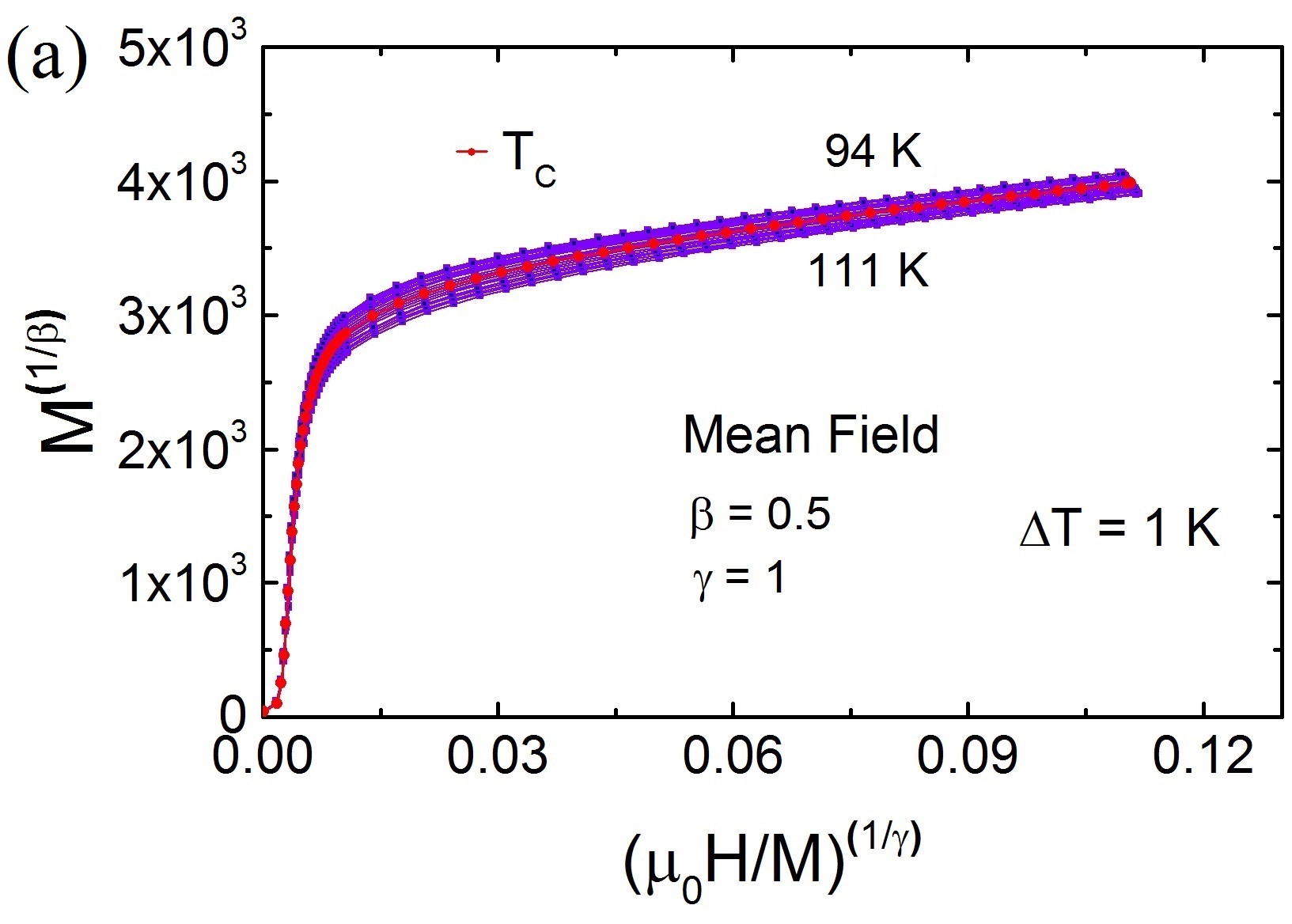}
		\includegraphics[trim=0.1mm 0.1mm 0.1mm 0.1mm,clip,width=85mm]{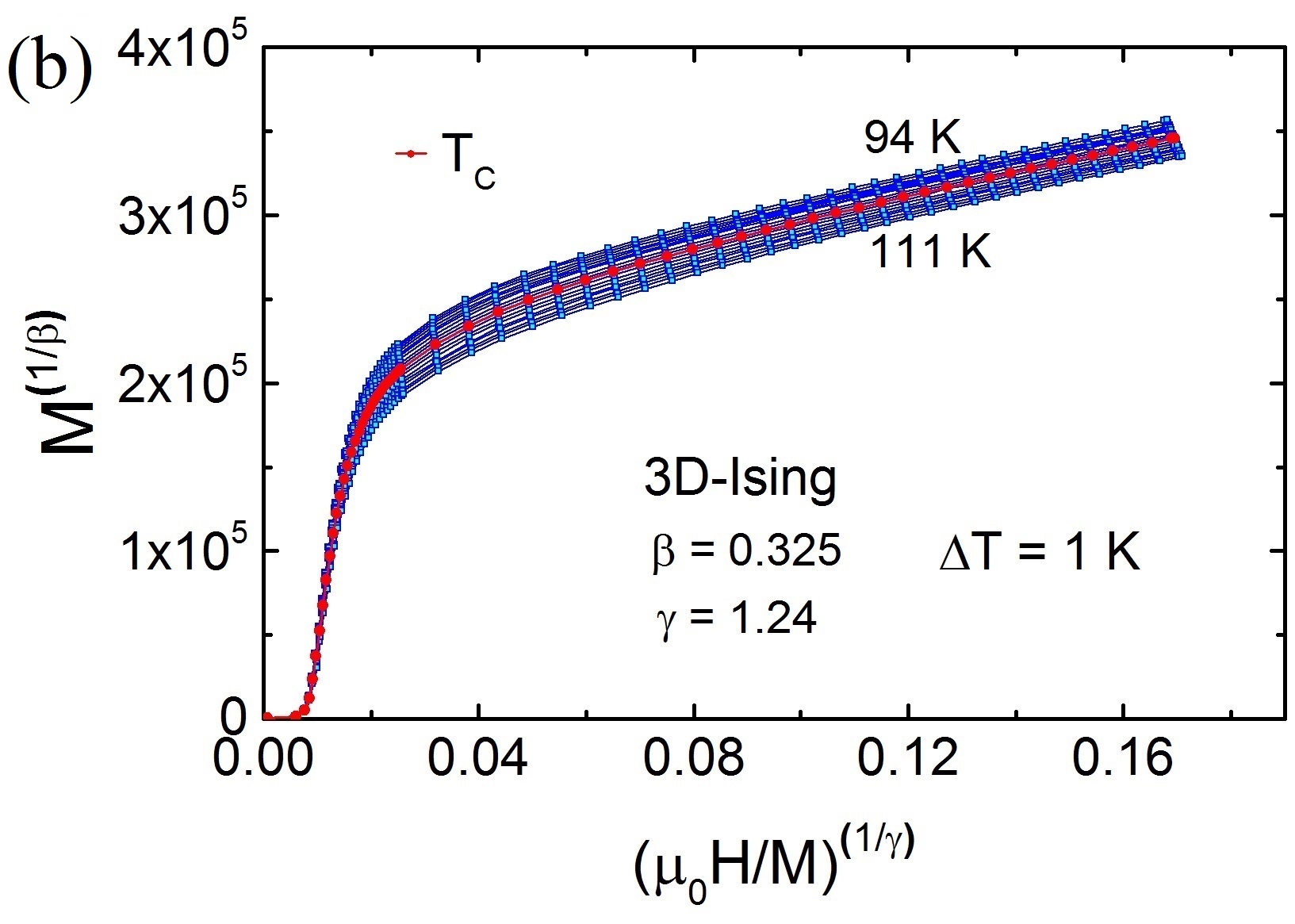}\\
		\includegraphics[trim=0.1mm 0.1mm 0.1mm 0.1mm,clip,width=85mm]{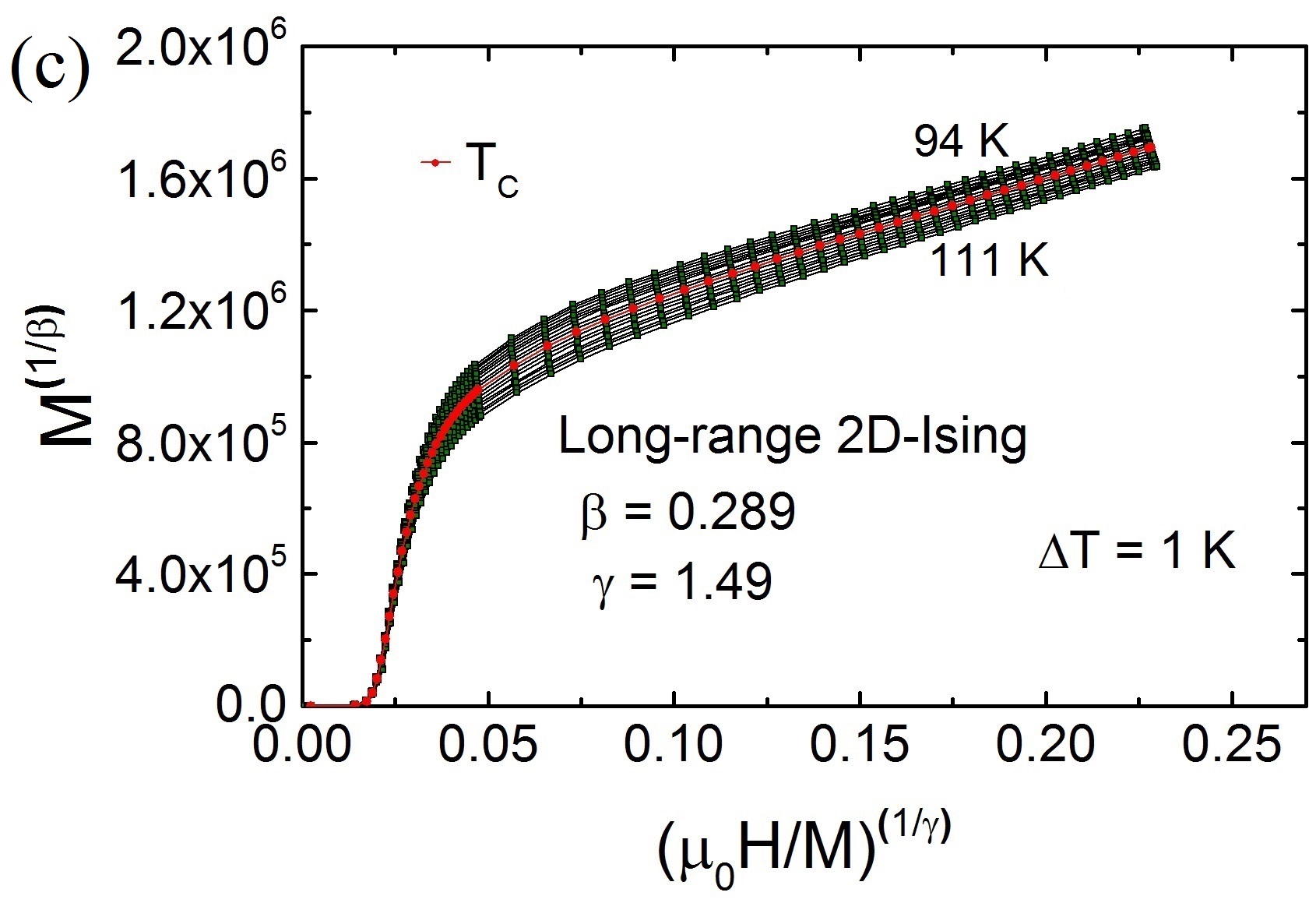}
		\includegraphics[trim=0.1mm 0.1mm 0.1mm 0.1mm,clip,width=85mm]{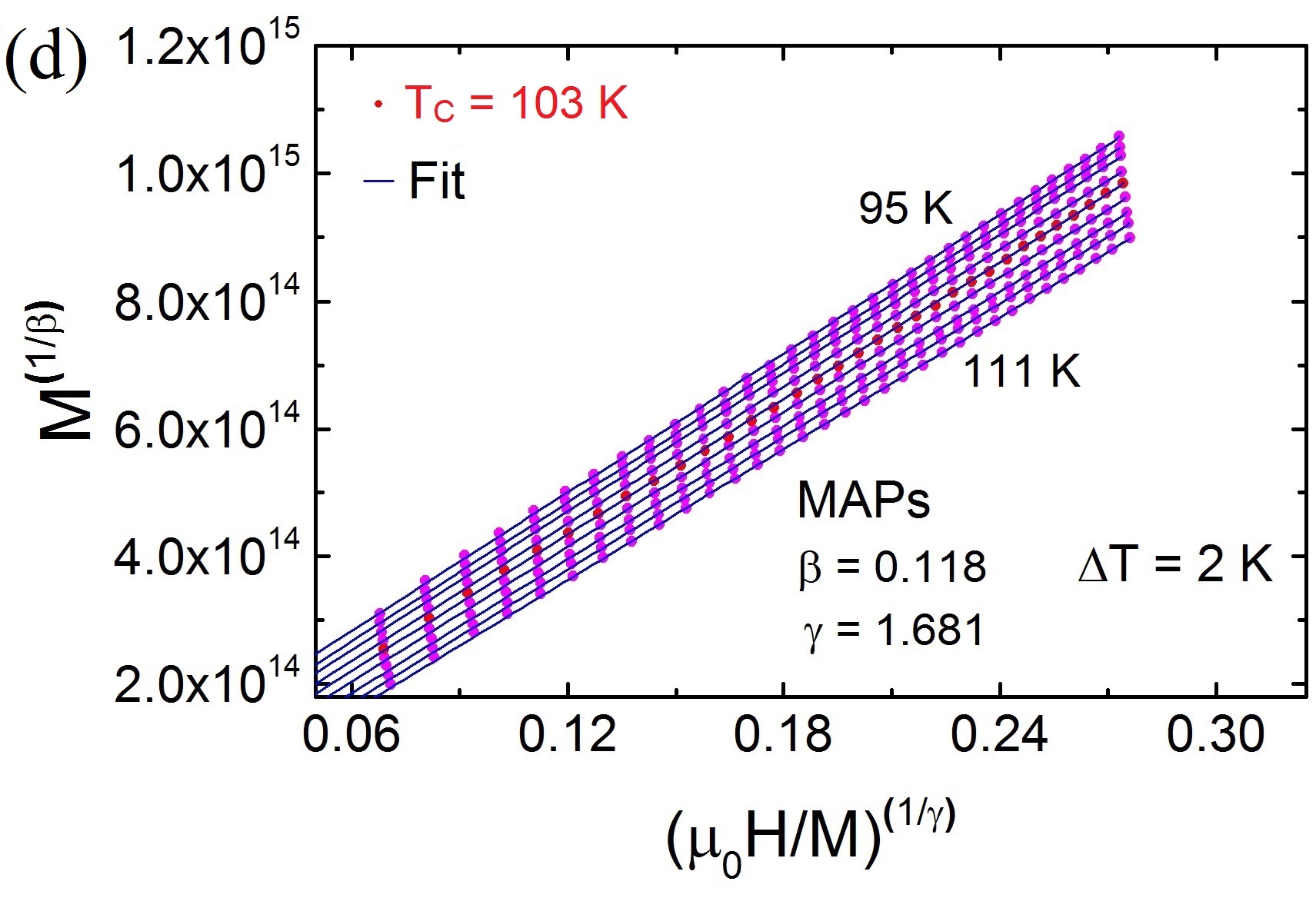}
	\end{tabular}
	\caption{Isotherms of M$^2$ vs. ${\mu{_0}}$H/M for 94 K $\leq$ T $\leq$ 111 K ($|\epsilon| \leq 0.1$) with (a) the Arrott plot (mean-field theory), which shows non-linear or non-mean-field type behavior even in higher magnetic field with positive slopes corresponding to the second-order phase transition. (b) MAPs for 3D-Ising with short-range interaction (c) MAPs for 2D-Ising model with Long-range interaction and (d) MAPs of M-H curves with $\beta$ = 0.118 and $\gamma$ = 1.681 in TL-LSMO-0.3. Solid lines are corresponding to the fits by the Eq. (\ref{Arrott Plot}).}
	\label{MAP}
\end{figure*}

\section{Critical analysis}

One way to determine the T$ _{C} $ and critical exponents ($\beta$ and $\gamma$) is Arrott analysis of the data\cite{domb2000phase}. Generally, if a magnetic system belongs to the mean-field ordering, i.e., $\beta$ = 0.5, $\gamma$ = 1, then Arrott plot (M$^2$ vs. ${\mu{_0}}$H/M) should results parallel lines and the isothermal curve at T$_C$ should pass through the origin\cite{domb2000phase}. The Arrot plot (M$^2$ vs. ${\mu{_0}}$H/M) for TL-LSMO-0.3 does not yield parallel lines around T$_C$, as shown in Fig. \ref{MAP}(a), which implies that there is non-mean-field type interaction in TL-LSMO-0.3. The positive slope in the Arrott plot confirms the second-order magnetic phase transition in TL-LSMO-0.3\cite{banerjee1964generalised}. Neither short-range 3D-Ising model ($\beta$ = 0.325, $\gamma$ = 1.24) nor long-range 2D-Ising model ($\beta$ = 0.289, $\gamma$ = 1.49) produce parallel lines, which are shown in Fig. \ref{MAP}(b) and \ref{MAP}(c). Therefore one can conclude that these two models cannot describe the critical behavior of TL-LSMO-0.3. Hence, we reanalyzed the magnetization isotherms of the TL-LSMO-0.3 by using the Arrott Noakes equation of state defined in the critical region, Eq. (\ref{Arrott Plot})\cite{arrott1967approximate}. The modified Arrott plots (MAPs) M$^{1/\beta}$ vs. ${\mu{_0}}$H/M$^{1/\gamma}$ for the M-H isotherms of TL-LSMO-0.3 in asymptotic region ($|\epsilon|<0.1$) is shown in Fig. \ref{MAP}(d). The value of the exponents $\beta$ and $\gamma$ are chosen such that the isotherms of MAPs display as close as parallel lines. The best fit of Eq. (\ref{Arrott Plot}) to the MAPs defined for TL-LSMO-0.3 in the temperature range 95 K $\leq$ T $\leq$ 111 K and field range 0.5 T $\leq$ $\mu_0$H $\leq$ 7 T yields the value of exponents $\beta$ = 0.118 $\pm$ 0.004, $\gamma$ = 1.681 $\pm$ 0.006 and T$_C$ = 103.54 $\pm$ 0.03 K.

\begin{figure}[htb!]
	\centering
	\includegraphics[trim=0.3mm 0.3mm 0.3mm 0.3mm,clip,width=\linewidth]{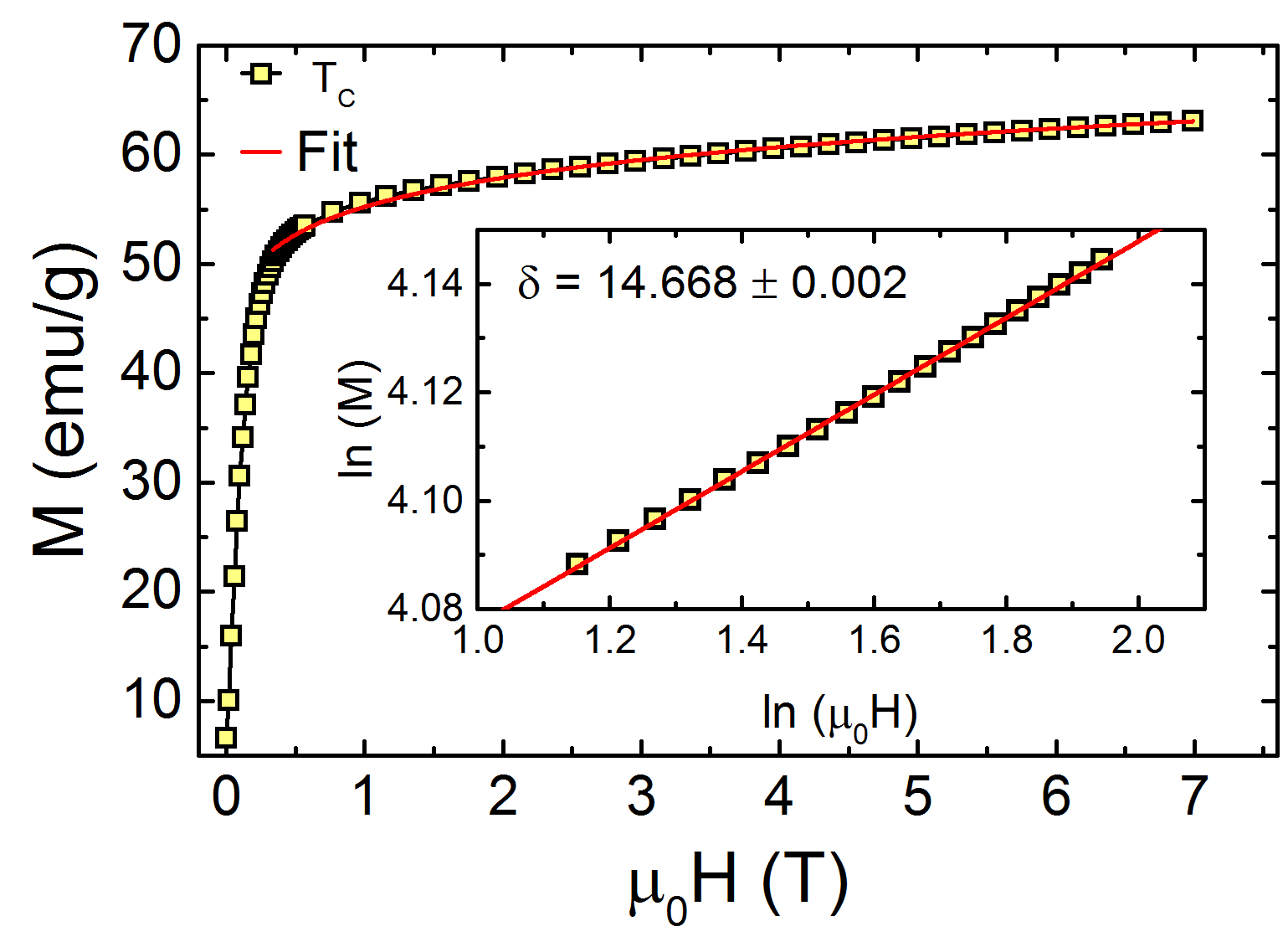}
	\caption{ M-H curve at T${_C}$ = 103 K and inset shows log-log plot of the curve. Solid line represents the fit to the Eq. (\ref{Del}) and yields $\delta$ = 14.668 $\pm$ 0.002 mentioned in graph is obtained from the fit to the Eq. (\ref{Del}). }
	\label{Delta}
\end{figure}

Next, we find out the value of exponent $\delta$ using M-H curve at T$_C$ and Eq. (\ref{Del}) as shown in Fig. \ref{Delta}. The value of the exponent $\delta$ = 14.668 $\pm$ 0.002 is obtained for TL-LSMO-0.3 by fitting the isotherm at T$_C$ to the Eq. (\ref{Del}). The value of the exponent $\delta$ for TL-LSMO-0.3 is larger than $\delta$ value in 3D universality classes defined for the short-range interaction, see Table \ref{Theory}. These exponents $\beta$, $\gamma$ and $\delta$ for TL-LSMO-0.3 should satisfy the Eq. (\ref{Widom})\cite{widom1965surface}. The value $\delta$ = 15.245 is obtained by using the value of $\beta$ and $\gamma$ determined from MAPS in Eq. (\ref{Widom}). The $\delta$ value obtained from Eq. (\ref{Widom}) is consistent with the $\delta$ value determined from the critical isotherm. Hence, both the exponents $\beta$ and $\gamma$ found to satisfy the Widom-scaling relation defined in Eq. (\ref{Widom}).

\begin{figure}[htb!]
	\centering
	\includegraphics[trim=0.3mm 0.3mm 0.3mm 0.3mm,clip,width=\linewidth]{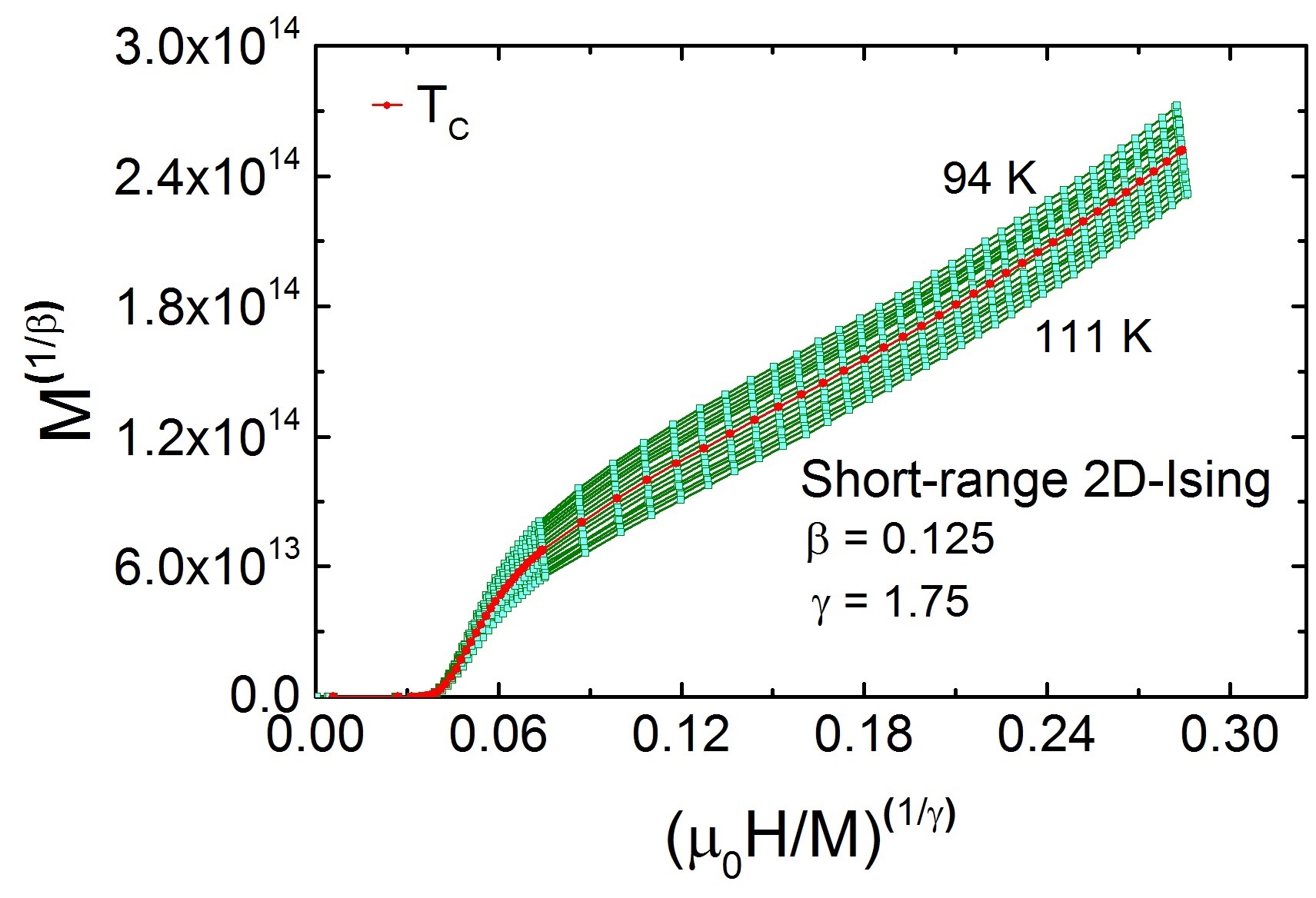}
	\caption{Isotherms of M$^2$ vs. ${\mu{_0}}$H/M for 94 K $\leq$ T $\leq$ 111 K ($|\epsilon| \leq 0.1$) with the short-range 2D-Ising model.}
	\label{2DSR}
\end{figure}

\begin{figure}[htb!]
	\centering
	\includegraphics[trim=0.3mm 0.3mm 0.3mm 0.3mm,clip,width=\linewidth]{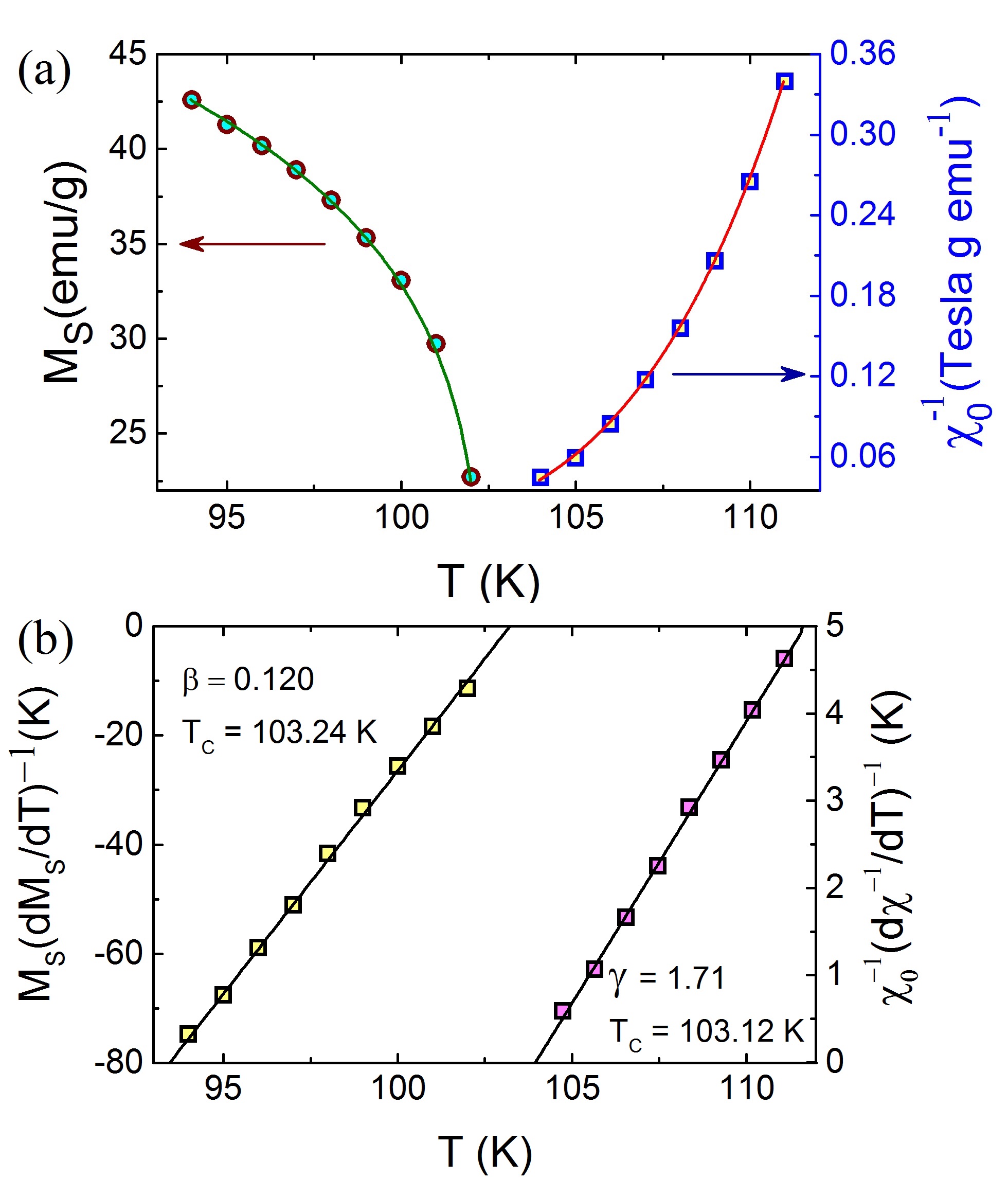}
	\caption{(a) Temperature variation of spontaneous magnetization M$_S$(T, 0) (left) and inverse susceptibility ${\chi}_0^{-1}$(T, 0) (right). (b) KF plots for the ${M_S}({dM_S(T)/dT})^{-1}$  (left) and ${\chi_0^{-1}}({d\chi_0^{-1}/dT})^{-1}$ (right).}
	\label{KFplot}
\end{figure}

Further, the exponents $\beta$ and $\gamma$ have been determined more accurately by Kouvel-Fisher (KF) method\cite{kouvel1964detailed}. The M$_S$ and $\chi_0$ are determined from the intersection with the axes M$^{1/\beta}$ and ${\mu{_0}}$H/M$^{1/\gamma}$, respectively. The intercepts are obtained from the linear extrapolation in the MAPs plotted for the 2D short-range Ising model because of the nearly parallel behavior of the isotherms as displayed in Fig. \ref{2DSR}. 
The variation of M$_S$ and $\chi_0$ with temperature for TL-LSMO-0.3 is shown in the Fig. \ref{KFplot}(a). The solid lines in the Fig. \ref{KFplot}(a) represent fit to the M$_S$ and $\chi_0$ using Eqs. (\ref{Ms}) and (\ref{chi}), respectively. The KF method has the following form, which is obtained by Eq. (\ref{Arrott Plot}) in the limit H $\rightarrow$ 0 for T $ < $ T$_C$ and T $ > $ T$_C$
\begin{equation}
\frac{M_S(T)}{dM_S(T)/dT} = \frac{T-T_C}{\beta}
\label{KFB}
\end{equation}
and
\begin{equation}
\frac{\chi_0^{-1}(T)}{d\chi_0^{-1}(T)/d(T)} = \frac{T-T_C}{\gamma}.
\label{KFG}
\end{equation}
The value of exponents $\beta$ and $\gamma$ can be determined from the slopes 1/$\beta$ and 1/$\gamma$ obtained from the linear variation of ${M_S}({dM_S(T)/dT})^{-1}$ vs. T and ${\chi_0^{-1}}({d\chi_0^{-1}/dT})^{-1}$ vs. T, respectively. The intersection with temperature axis yields T$_C$ as shown in the Fig. \ref{KFplot}(b). Solid lines in Fig. \ref{KFplot}(b) represent the fit to the ${M_S}({dM_S(T)/dT})^{-1}$ vs. T and ${\chi_0^{-1}}({d\chi_0^{-1}/dT})^{-1}$ vs. T using Eqs. (\ref{KFB}) and (\ref{KFG}), respectively. The KF method results the value of exponents $\beta$ = 0.120 $\pm$ 0.003 with T$_C$ = 103.24 $\pm$ 0.01 K and $\gamma$ = 1.710 $\pm$ 0.005 with T$_C$ = 103.12 K $\pm$ 0.02. These results are consistent with the value of exponents obtained from the MAPs.
  
\begin{figure}[htb!]
	\centering
	\includegraphics [trim=0.3mm 0.3mm 0.3mm 0.3mm,clip,width=\linewidth]{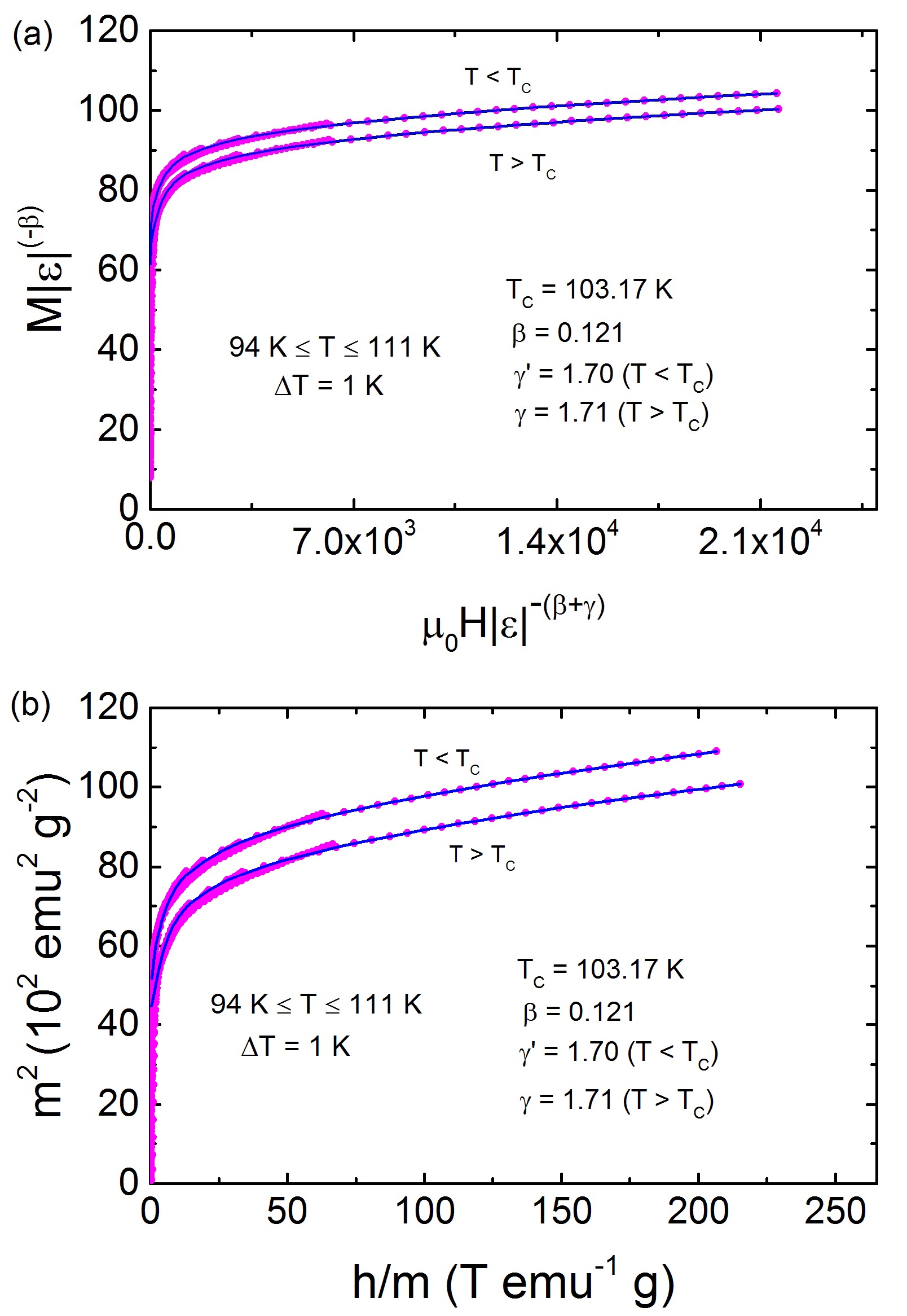}
	\caption{(a) Scaling of the M-H curve below and above T$_C$ for TL-LSMO-0.3 in critical region $\epsilon$ $\leq$ 0.1 using Eq. (\ref{scal}). Solid lines show best fit polynomials. Collapse of all the M-H curve on single curve below and above T$_C$ separately confirms the validity of obtained exponents. (b)  m$^2$ vs h/m below and above $T_C$ also shows collapse of curve again validate the obtained results.}
	\label{Scaling}
\end{figure}

Further, we confirm that the obtained exponents are not different above and below T$_C$. As we know that, one can also deduce the critical exponents of a magnetic sample using scaling theory, which states that for an appropriate value of the critical exponents ($\beta$ and $\gamma$), the plot of scaled magnetization (M${\epsilon}^{-\beta}$) vs. renormalized field ($\mu_0$H$\epsilon^{-(\beta+\gamma)}$) should fall onto two separate curves: one for T $ < $ T$_C$ and other for T $ > $ T$_C$. Figure \ref{Scaling} represents the M${\epsilon}^{-\beta}$ as a function of $\mu_0$H$\epsilon^{-(\beta+\gamma)}$ below and above T$_C$ in TL-LSMO-0.3. One can see that all the magnetization curve fall onto two curves below and above T$_C$ separately, when the value of T$_C$ and exponents are chosen as T$_C$ = 103.17 $\pm$ 0.01 K, $\beta$ = 0.121 $\pm$ 0.001, {$\gamma$}$^{'}$ = 1.710 $\pm$ 0.005 for T $ < $ T$_C$ and {$\gamma$} = 1.702 $\pm$ 0.002 for T $ > $ T$_C$ in Eq (\ref{scal}). Further, we have plotted m$^2$ vs. h/m and again found that all the data collapse onto two separate curves above and below T$ _{C} $, respectively. This confirms that the critical exponents are reliable, unambiguous and the interactions get renormalized appropriately following the scaling equation of state in the critical regime.
\begin{table*}[htb] 
	\caption{Comparison of critical exponents $\beta$, $\gamma$ and $\delta$ of TL-LSMO-0.3 with various theoretical models for three dimension and two dimension. RG: Renormalization group, CI: Critical isotherm. }
	\begin{ruledtabular}
		\begin{tabular}{ccccccc}
			{} & Method & {$T_C$}(K)& $\alpha$ & {$\beta$} & {$\gamma$} & {$\delta$}\\ \hline
			(Theory)  \\
			Mean Field~\cite{kim2002critical, kadanoff1966scaling} & {}&{}&0& 0.5 & 1 & 3  \\
			Tricritical Mean Field~\cite{kim2002critical, huang1987statistical}&{}&{}&0&0.25&1&5\\
			3D-ISing (d=3, n=1) ~\cite{kim2002critical, kadanoff1966scaling} &{RG-$\phi$$^{4}$}&{}& 0.11&0.325 & 1.241 & 4.82 \\
			3D-XY (d=3, n=2)~ \cite{kim2002critical, kadanoff1966scaling} & {RG-$\phi$$^{4}$} &{}&{-0.007}& 0.346 & 1.316 & 4.81 \\
			3D-Heisenberg (d=3, n=3)~\cite{kim2002critical, kadanoff1966scaling}&{RG-$\phi$$^{4}$} &{} &{-0.115}&0.365 & 1.386 & 4.8 \\
			Short-range 2D-ISing~\cite{domb2000phase, fisher1974renormalization} &{Onsager solution}&{}&{0}& 0.125 & 1.75 & 15 \\
			Long-range 2D-ISing~\cite{fisher1974renormalization} &{RG-$\epsilon$$^{'}$}&{}&{}& 0.289 & 1.49 & 6 \\
			\hline
			(Experiment)\\
			{La$_{2.1}$Sr$_{1.9}$Mn$_{3}$O$_{10}$}[This work] &MAPs& 103.54 $\pm$ 0.03 &{}&0.118 $\pm$ 0.004&{1.681 $\pm$ 0.006}&\\
			{}&CI&{}&{}&{}&{}&14.668 $\pm$ 0.002\\
			{}&KF&103.24 $\pm$ 0.01&{}&0.120 $\pm$ 0.003&\\
			{}&{}&{103.12 $\pm$ 0.02}&{}&{}&1.710 $\pm$ 0.005&\\
			{}&{RG}&{}&{}&{0.145}&{1.91}&{14.172}\\
			{}&{Scaling}&{103.17 $\pm$ 0.01}&{}&{0.121 $\pm$ 0.001}&{1.710 $\pm$ 0.005}&{}
		\end{tabular}
	\end{ruledtabular}
	\label{Theory}
\end{table*} 

\section{spin interaction}

Finally, we have discuss the range and dimensionality of the TL-LSMO-0.3 with the help of the renormalization group theory. For a homogeneous magnet the universality class of the magnetic phase transition is defined by the interaction J(r). Fisher \textit{et al.}\cite{fisher1974renormalization} used renormalization group theory and suggested that the exchange interaction decays with distance r as J(r) $\sim {r^{-(d+\sigma)}}$, where d is dimensionality and $\sigma$ is the range of the interaction. Also, they have discussed the validity of such a model for $\sigma$ $ < $ 2 having long-range interactions. Further, the critical exponent $\gamma$ associated with the susceptibility can be given as
\begin{eqnarray}
\begin{split}
 \gamma = 1 + \frac{4}{d}\Bigg(\frac{n+2}{n+8}\Bigg)\Delta{\sigma}+\Bigg(\frac{8(n+2)(n-4)}{d^2(n+8)^2}\Bigg) \nonumber\\ \times \Bigg[{1+\frac{2G(\frac{d}{2})(7n+20)}{(n-4)(n+8)}}\Bigg]\Delta{\sigma^2},
 \label{spin}
\end{split}
\end{eqnarray}
where $\Delta{\sigma}$ = $\Big(\sigma - {\frac{d}{2}}\Big)$ and $G(\frac{d}{2}) = 3$ - $\frac{1}{4}$$\big(\frac{d}{2}\big)^2$. The range of interaction $\sigma$ and dimensionality of both space and spin, is determined by the same procedure defined in the Ref \cite{fischer2002critical}. The value of exchange interaction $\sigma$ is chosen for a particular set of $\left\{d:n\right\}$ such that the Eq. (\ref{spin}) results the value of exponent $\gamma$ close to the experimentally determined, $\gamma$ = 1.71. Further, the remaining exponents can be determined with the help of Eqs. (\ref{Rushbrooke}), (\ref{Widom}) and $\sigma$ value using following expressions: $\alpha$ = 2- $\nu$d, $\nu$ = $\gamma$/$\sigma$, $\gamma$ = $\nu$(2 - $\eta$) and $\eta$ = 2-$\sigma$. We found that $\left\{d:n\right\}$ = $\left\{2:1\right\}$ yields a value $\sigma$ = 1.69. The value of $\sigma$ = 1.69 is then used to determine the remaining exponents; $\beta$ = 0.135, $\gamma$ = 1.91 and $\delta$ = 14.172, which are close to the value of exponents obtained from previous methods MAPs, KF and scaling analysis (Table \ref{Theory}). We have also examined the remaining 3D and 2D models but they cannot describe the experimental results obtained for TL-LSMO-0.3. For example, the 3D-Heisenberg $\left\{d:n\right\}$ = $\left\{3:3\right\}$, 3D-XY $\left\{d:n\right\}$ = $\left\{3:2\right\}$ and 3D-Ising models $\left\{d:n\right\}$ = $\left\{3:1\right\}$ with short-range exchange interaction yield the value of exponent $\gamma$ = 1.25, 1.27 and 1.23, respectively. Similarly, the 2D Heisenberg $\left\{d:n\right\}$ = $\left\{2:3\right\}$ and the 2D XY $\left\{d:n\right\}$ = $\left\{2:2\right\}$ models defined for the short-range exchange interaction yield $\gamma$ = 2.56 and 2.30, respectively. The other calculated exponents ($\beta$ and $\delta$) by using respective $\sigma$ values for different models $\left\{d:n\right\}$ = $\left\{2:2, 3\right\}$, $\left\{d:n\right\}$ = $\left\{3:1, 2, 3\right\}$ also show significant difference from experimental results for $\beta$ and $\delta$. Hence, all the other 2D and 3D models can be discarded. The long-range mean field model is valid for $\sigma$ $\leq$ 3/2 and j(r) decreases as j(r) $\sim$ r$^{-4.5}$. For $\sigma$ $\geq$ 2 only short-range 3D-Heisenberg model is valid and j(r) varies as j(r) $\sim$ r$^{-5}$. The other 3D universality classes for short-range lies between 3/2 $<$ $\sigma$ $<$ 2, where j(r) decreases as j(r) $\sim$ r$^{-d-\sigma}$.  All the theoretical models with short-range exchange interaction varies with distance r as J(r) $\sim$ e$^{-(r/b)}$ (where b is correlation length). The renormalization group analysis suggests that the spin interaction in TL-LSMO-0.3 is of a short-range 2D Ising $\left\{d:n\right\}$ = $\left\{2:1\right\}$ type with $\sigma$ = 1.69 and decays as $\sim$ r$^{-3.69}$.

\section{Discussion} 

\begin{figure}[htb!]
	\centering
	\includegraphics[trim=0.3mm 0.3mm 0.3mm 0.3mm,clip,width=\linewidth]{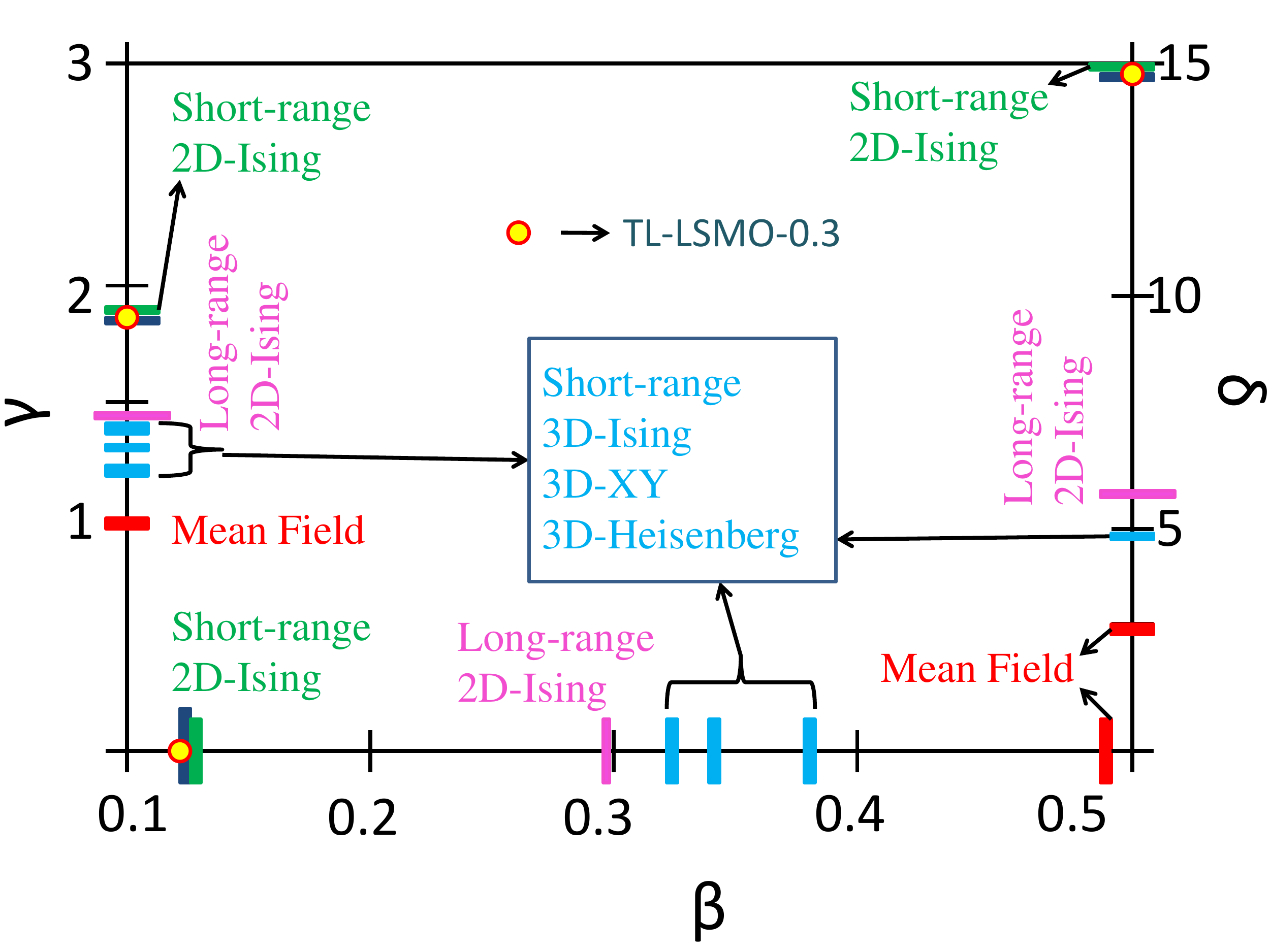}
	\caption{Comparison of different exponents of TL-LSMO-0.3 (denoted as closed circles filled with yellow color) with that of standard universality classes. Different vertical bars represents different models for short-range and long-range exchange interactions. Short-range $\rightarrow$ the 2D-Ising (green bar), 3D-Ising (sky blue bar), 3D-XY (sky blue bar) and 3D-Heisenberg (sky blue bar) and long-range $\rightarrow$ 2D-Ising model (pink bar).}
	\label{Models}
\end{figure}

All the findings in the above sections for TL-LSMO-0.3 yield the value of critical exponents close to the short-range 2D-Ising model. A graphical comparison of the critical exponents $\beta$, $\gamma$ and $\delta$ for TL-LSMO-0.3 with the various theoretical models is represented in Fig. \ref{Models}. The obtained results of the exponents consistent with the Q2D-layered structural characteristic of TL-LSMO-0.3 and emphasize that the magnetic anisotropy is playing a crucial role in the magnetism of the TL-LSMO-0.3.
The 2D magnetism in TL-LSMO-0.3 also emphasizes that the inter-layer interaction is weakened around T$_C$. In contrast, the intra-layer interaction becomes stronger, which leads to a 2D FM in TL-LSMO-0.3. Our results for TL-LSMO-0.3 are consistent with the Taroni \textit{et al.}\cite{taroni2008universal} criterion, according to which the value of critical exponent $\beta$ for 2D magnets should lie in the 0.1 $\geq$ $\beta$ $\geq$ 0.25. Similar results have been reported in bi-layer La$_{2-2x}$Sr$_{1+2x}$Mn$_{2}$O$_{7}$ in which short-range 2D FM ordering occurs around T$_C$\cite{osborn1998neutron, gordon1999specific}. Osborn \textit{et al.}\cite{osborn1998neutron} performed neutron scattering measurement in bi-layer La$_{2-2x}$Sr$_{1+2x}$Mn$_{2}$O$_{7}$ for x = 0.4 and claimed that, there is a short-range 2D-Ising interaction with $\beta$ = 0.13 $\pm$ 0.01. Gordon \textit{et al.}\cite{gordon1999specific} performed specific heat measurement on La$_{2-2x}$Sr$_{1+2x}$Mn$_{2}$O$_{7}$ for x = 0.4 and claimed that the obtained result is consistent with 2D-XY or 2D-Ising critical fluctuation. There is no neutron diffraction data on tri-layer La$ _{3-3x} $Sr$ _{1+3x} $Mn$ _{3} $O$ _{10} $. However, one can get an idea about the spin structure and spin-spin interaction from the neutron diffraction data for bi-layer manganites.

Next, we discuss the unconventional behavior of temperature-dependent magnetization and magnetic spin structure in different regions. Conventionally, when an FM material undergoes a magnetic phase transition from FM to PM sate, the magnetic moment of the system becomes zero above T$ _{C} $. As shown in Fig. \ref{FC_ZFC}, the magnetic moment of the FC curve for TL-LSMO-0.3 is non-zero even above T$ _{C} $ and also another transition appears at higher temperature $\approx$ 263 K, denoted as T$ ^{*} $. The non-zero magnetization above T$ _{C} $ emphasizes that the phase transition in TL-LSMO-0.3 is not an FM to PM state.
\begin{figure}[htb!]
	\centering
	\includegraphics[trim=0.3mm 0.3mm 0.3mm 0.3mm,clip,width=\linewidth]{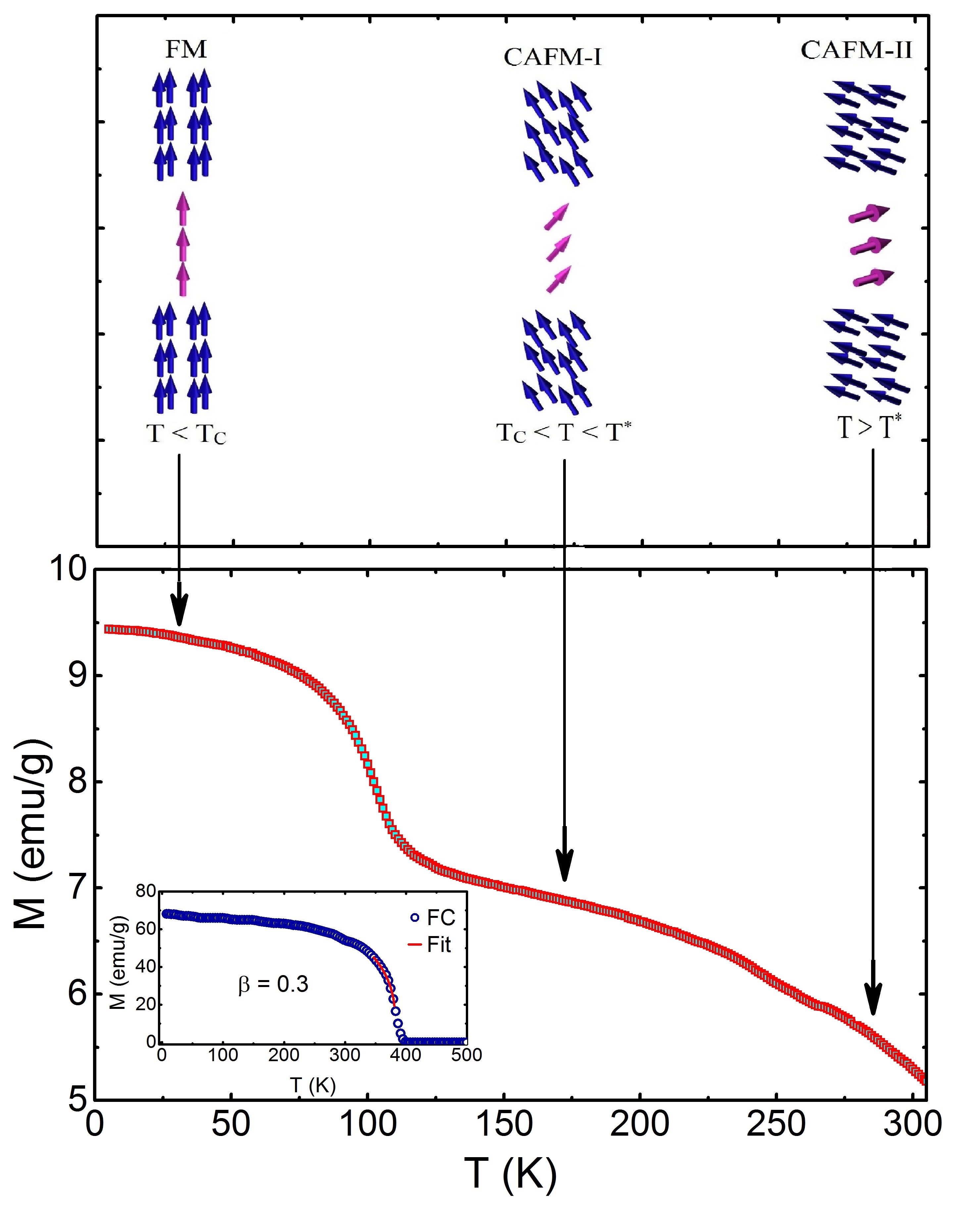}
	\caption{ Spin structure in the different temperature range of TL-LSMO-0.3. Below T$ _{C} $, i.e., T $ <  $ T$ _{C} $ the spins have FM alignment. First type of canting of spins occur in the range T$ _{C} $ $ < $ T $ < $ T$ ^{*} $ and above T$ ^{*} $ second type of canting is represented. The inset shows the FC curve for infinite-layer La$ _{0.7} $Sr$ _{0.3} $MnO$ _{3} $\cite{shi1999electrical}. The solid line in red color represents the fit to the Eq. \ref{Ms} and yield the value of exponent $\beta$ = 0.3. }
	\label{canted}
\end{figure}
A similar magnetic phase transition has been observed in bi-layer La$_{2-2x}$Sr$_{1+2x}$Mn$_{2}$O$_{7}$ RP series manganites and extensive studies have been conducted to explore the magnetic structure of this region (between T$ _{C} $ and T$ ^{*} $). Kimura \textit{et al.}\cite{kimura1996interplane} studied La$_{1.4}$Sr$_{1.6}$Mn$_{2}$O$_{7}$ and claimed that there is a short-range 2D FM ordering between T$ _{C} $ and T$ ^{*} $, which give rise the finite magnetic moment in this region and above T$ ^{*} $ system goes to the PM state. The disagreement of 2D FM characteristic above T$ _{C} $ was shown by Heffner \textit{et al.}\cite{heffner1998effects} from the muon spin rotation study in bi-layer La$_{1.4}$Sr$_{1.6}$Mn$_{2}$O$_{7}$ and claimed that there is no evidence of 2D magnetic ordering above T$ _{C} $. Later, a rigorous neutron scattering study in bi-layer La$_{2-2x}$Sr$_{1+2x}$Mn$_{2}$O$_{7}$ for x = 0.4 by Osborn \textit{et al.}\cite{osborn1998neutron} revealed that there is a strong canting of the Mn spins in adjacent MnO$ _{2} $ layers within each  MnO$ _{2} $ bi-layer above T$ _{C} $ and the canting angle depends on both temperature and magnetic field. Their neutron scattering study revealed that the FM and antiferromagnetic (AFM) magnetic ordering are inhomogeneously distributed in approximately equal volume above T$ _{C} $. Therefore, a non-collinear spin correlation or the canting of spins arises due to the competing FM DE and AFM superexchange (SE) interaction. The intra-bi-layer or inter-planer interaction is substantially weaker than the intra-planar interaction, which produces a large magnetic anisotropy in the exchange interactions. The FM in manganites is governed by the DE mechanism based on the hopping of the electrons and the kinetic energy of mobile electrons is lowered by polarizing the Mn spins, which are localized in the MnO$ _{2} $ plane. Hence, there would be larger free energy within the plane due to the energy acquired from delocalized electrons\cite{osborn1998neutron}. A comparatively large number of Mn spins can take part in the FM cluster than among the adjacent planes, where only two Mn spin sites are available. Therefore, the SE interaction strongly affects the spin interactions along the c-axis compared to the spins in the ab-plane\cite{osborn1998neutron}. This is the reason why intra-planar interaction is stronger than intra-bi-layer interaction. 

Based on the above discussions for non-zero magnetization in bi-layer La$_{2-2x}$Sr$_{1+2x}$Mn$_{2}$O$_{7}$,  we propose that a similar inhomogeneous distribution of FM and AFM clusters are giving rise the canted AFM-I (CAFM-I) type spin structure above T$ _{C} $, which is responsible for the non-zero magnetization in TL-LSMO-0.3. The magnetic moment of the system is continuously decreasing with increase in temperature and the system undergoes another transition at T$ ^{*} $. It is noted that the magnetic moment is still non-zero above second transition T$ ^{*} $, which indicate that the system is going to another canted AFM-II (CAFM-II) state with a canting angle greater than that of the CAFM-I state and responsible for a finite magnetic moment above the second transition T$ ^{*} $. The layered crystal structure of the TL-LSMO-0.3 suggests that there are different type of interactions present in the system; inter-tri-layer ($ J' $), inter-layer or intra-tri-layer ($ J_{c} $) and intra-planer ($ J_{ab} $) interaction. The two manganese ions Mn$ ^{3+} $ and Mn$ ^{4+} $ are distributed in the ratio of Mn$ ^{3+} $:Mn$ ^{4+} $ 2.33:1 in the system. Depending upon the distribution and distance between Mn spins in different directions the order of the strength of interaction can be given as $ J_{ab} $ $ > $ $ J_{c} $ $ >>  $$ J' $. The intra-planer interaction $ J_{ab} $ is the DE interaction, i.e., the spins are coupled ferromagnetically and the intra-tri-layer interaction $ J_{c} $ is SE interaction, which implies that the spins are coupled antiferromagnetically. On the other hand, the inter-tri-layer interaction $ J' $ is the direct exchange interaction. Although the DE interaction is strong in ab-plane, but a minor AFM interaction also coexists. Similarly, the intra-tri-layer interaction $ J_{c} $ is an SE interaction along the c-axis, but a weak DE interaction can also coexist. The intra-planer interaction $ J_{ab} $ is the strongest and dominates over other interactions ($ J_{c} $ and $ J' $). $ J_{ab} $ combined with the anisotropy give rise to the 2D-Ising like spin structure below T$ _{C} $. With an increase in temperature, the weakest interaction $ J' $ breaks at first transition T$ _{C} $ and give rise to the phase transition from 2D-Ising to the CAFM-I state. The relative strength of intra-planer interaction $ J_{ab} $ decreases with increase in temperature and hence the intra-tri-layer interaction $ J_{c} $ started competing with $ J_{ab} $. The competition between these two magnetic interactions results in the CAFM-I state. The intra-tri-layer minor FM interaction along c-axis and intra-layer AFM interaction in ab-plane are weakened with further increase in temperature resulting in a phase transition from CAFM-I to CAFM-II at T$ ^{*} $. The CAFM-II state is observed due to the competition between SE exchange interaction along the c-axis and DE interaction in the ab-plane. Figure \ref{canted} shows the possible magnetic structures above and below T$ _{C} $ in TL-LSMO-0.3 based on the bi-layer studies\cite{sonomura2013neutron}. The inset of Fig. \ref{canted} shows the FC curve for infinite-layer La$ _{0.7} $Sr$ _{0.3} $MnO$ _{3} $\cite{shi1999electrical}. The solid line in red color represents the fit to the Eq. \ref{Ms} and yield the value of exponent $\beta$ = 0.3, which is close to the 3D-Ising universality class. Michael \textit{et al.}\cite{martin1996magnetism} and Vasiliu \textit{et al.}\cite{vasiliu1998spin} performed neutron scattering in infinite-layer La$ _{0.7} $Sr$ _{0.3} $MnO$ _{3} $ and shown that it belongs to the short-range 3D-Ising universality class with $\beta$ = 0.295 and 0.3, respectively. One can see that the magnetic moment of infinite-layer La$ _{0.7} $Sr$ _{0.3} $MnO$ _{3} $ above T$ _{C} $ is zero in contrast to TL-LSMO-0.3. With reduced dimensionality from 3D to Q2D, the system changes from 3D-Ising to 2D-Ising like spin-spin interaction and there exists a canted AFM magnetic structure between FM and PM state due to the existence of different exchange interactions. It is well known that the different interactions are responsible for different spin structures. The exchange interaction aligns the spins parallel to each other. In contrast, long-range dipolar interaction favors a close loop of spins and anisotropy energy favors perpendicular alignment of spins to the plane. Hence, the anisotropy in a magnetic system results in the Ising spin structure and the system behaves as a uniaxial magnet. The Ising interaction below T$ _{C} $ in TL-LSMO-0.3 emphasizes that magnetic anisotropy plays a crucial role in the magnetism of the TL-LSMO-0.3. It is believed that the skyrmions in manganite perovskites result from the competition between different energies such as exchange interaction, long-range dipolar interaction and anisotropy energy. Keeping in view of the observation of bi-skyrmion in the bi-layer La$_{1.37}$Sr$_{1.63}$Mn$_{2}$O$_{7}$, which has similar magnetic properties to the TL-LSMO-0.3, we contemplate that TL-LSMO-0.3 should also host the skyrmions. 

All the above discussions and experimental observations imply that much more experimental and theoretical works are needed to thoroughly understand the magnetism in tri-layer La$_{3-3x}$Sr$_{1+3x}$Mn$_{3}$O$_{10}$ manganite perovskite. The magnetic and transport properties of tri-layer La$_{3-3x}$Sr$_{1+3x}$Mn$_{3}$O$_{10}$ manganite perovskite for different Sr concentration is not yet explored. Therefore, it is highly desirable to establish the structural, magnetic and electronic phase diagram of tri-layer La$_{3-3x}$Sr$_{1+3x}$Mn$_{3}$O$_{10}$ because it may be a potential candidate for the future spintronics. We hope the present study will prompt further investigation in understanding the magnetic phase transition and different types of exchange interaction in the low dimensional RP series manganite perovskites. 

\section{Conclusion}

In summary, we have established an understanding of the phase transition in a novel quasi-two-dimensional ferromagnetic tri-layer La$_{2.1}$Sr$_{1.9}$Mn$_{3}$O$_{10}$ RP series manganite. We have discussed the low dimensionality in the magnetic properties of the tri-layer La$_{2.1}$Sr$_{1.9}$Mn$_{3}$O$_{10}$ manganite perovskite. A comprehensive experimental study of the critical properties is performed using isothermal magnetization in the vicinity of the Curie temperature T$_C$. We have used various techniques, including the modified Arrott plots (MAPs), Kouvel-Fisher (KF) method, scaling and critical isotherm analysis to determine the critical exponents of the La$_{2.1}$Sr$_{1.9}$Mn$_{3}$O$_{10}$. The obtained critical exponents for La$_{2.1}$Sr$_{1.9}$Mn$_{3}$O$_{10}$ are close to theoretical values compatible with 2D-Ising model with short-range interaction. The critical exponents of the La$_{2.1}$Sr$_{1.9}$Mn$_{3}$O$_{10}$ were also determined by using renormalization group approach for a two-dimensional (2D) Ising system with short-range interactions decaying as j(r) $\sim$ r$^{-d-\sigma}$ with $\sigma$ = 1.69. We suggest that the strong anisotropy and layered structure are playing a crucial role resulting Ising-like interaction in La$_{2.1}$Sr$_{1.9}$Mn$_{3}$O$_{10}$. Based on results obtained for La$_{2.1}$Sr$_{1.9}$Mn$_{3}$O$_{10}$ in the present study, we propose that the La$_{3-3x}$Sr$_{1+3x}$Mn$_{3}$O$_{10}$ can be a potential candidate for the skyrmion host material. Finally, we propose that the non-zero magnetic moment above T$ _{C} $ is due to the canted antiferromagnetic spin orientation.

\bibliographystyle{apsrev4-2}
\bibliography{Ref}
\end{document}